 \documentclass[preprint,3p,times,review]{elsarticle}

\usepackage{amssymb}
\usepackage[utf8]{inputenc}
\usepackage{amsmath}
\usepackage{graphicx}
\usepackage{siunitx}
\usepackage[export]{adjustbox}
\usepackage{subfig}
\usepackage{framed}
\usepackage{color,soul}
\usepackage[flushleft]{threeparttable}
\usepackage{caption}
\usepackage{floatrow}
\usepackage{multirow}
\usepackage{booktabs}
\usepackage{tabularx}
\usepackage{hyperref}
\usepackage{xcolor}   
\usepackage{color, soul}

\DeclareUnicodeCharacter{2212}{-}
\DeclareUnicodeCharacter{0308}{-}
\DeclareUnicodeCharacter{0001}{-}
\DeclareUnicodeCharacter{200E}{-}
\def\etal{{\it \etal. }}

	\def\etal{{\it et al. }}

\journal{PCCP}

\begin{document}

\begin{frontmatter}

\title{Navigating the Evolution of Two-dimensional Carbon Nitride Research: Integrating Machine Learning into Conventional Approaches}
\author[1]{Deep Mondal}
\author[2]{Sujoy Datta\corref{ac}} 
\ead{sujoydatta13@gmail.com}
\cortext[]{DM and SD contributed equally to this work}
\author[1]{Debnarayan Jana\corref{ac}}
\ead{djphy@caluniv.ac.in}
\cortext[ac]{Corresponding author}
\address[1]{Department of Physics, University of Calcutta, 92 A. P. C. Road, Kolkata-700009, India}
\address[2]{Kadihati KNM High School, P.O. Ganti, Kolkata-700132, India}

\begin{abstract}
Carbon nitride research has reached a promising point in today's research endeavours with diverse applications including photocatalysis, energy storage, and sensing due to their unique electronic and structural properties. Recent advances in machine learning (ML) have opened new avenues for exploring and optimizing the potential of these materials. This study presents a comprehensive review of the integration of ML techniques in carbon nitride research with an introduction to CN classifications and recent advancements. We discuss the methodologies employed, such as supervised learning, unsupervised learning, and reinforcement learning, in predicting material properties, optimizing synthesis conditions, and enhancing performance metrics. Key findings indicate that ML algorithms can significantly reduce experimental trial-and-error, accelerate discovery processes, and provide deeper insights into the structure-property relationships of carbon nitride. The synergistic effect of combining ML with traditional experimental approaches is highlighted, showcasing studies where ML driven models have successfully predicted novel carbon nitride compositions with enhanced functional properties. Future directions in this field are also proposed, emphasizing the need for high-quality datasets, advanced ML models, and interdisciplinary collaborations to fully realize the potential of carbon nitride materials in next-generation technologies. 
\end{abstract}
\end{frontmatter}

\begin{table}[htbp]
	\centering
	\caption{\textbf{List of Abbreviations}}
	\renewcommand{\arraystretch}{1.1} 
	\setlength{\tabcolsep}{3pt} 
	\begin{tabular}{ll ll} 
		\hline
		\hline
		\textbf{Abbreviation} & \textbf{Definition} & \textbf{Abbreviation} & \textbf{Definition} \\
		\hline
		2D  & Two-dimensional materials  & MLIP  & Machine Learning Interatomic Potential \\
		CN  & Carbon Nitrides & ELM  & Extreme Machine Learning \\
		PBE & Perdew-Burke-Ernzerhof & HSE  & Heyd-Scuseria-Ernzerhof \\
		DFT  & Density Functional Theory & DOS  & Density Of States \\
		SCAN  & Strongly Constrained and Appropriately Normed & ILs  & Ionic Liquids \\
		vdW  & van der Waals & TEM & Transmission Electron Microscopy \\
		XPS & X-ray Photoelectron Spectroscopy & STS & Scanning Tunneling Spectroscopy \\
		PPL & Poly-Propio-Lactone & USPEX & Universal Structure Predictor \\
		GNN & Graph Neural Network & KNR & k-Nearest Neighbors Regression \\
		(K)RR & (Kernel) Ridge Regression & GBR & Gradient Boosting Regression \\
		XGBoost & Extreme Gradient Boosting Regression & SVR & Support Vector Regression \\
		LASSO & Least Absolute Shrinkage and Selection Operator & RF & Random Forest Strategy \\
		SVM & Support Vector Machine & GPR & Gaussian Process Regression \\
		ReLU & Rectified Linear Unit & PCA & Principal Component Analysis \\
		AIMD & Ab-Initio Molecular Dynamics & NNP & Neural Network Potential \\
		SAC & Single Atom Catalysts & TM & Transition Metals \\
		SISSO & Sure Independence Screening & PSO & Particle Swarm Optimization \\
		ECL & Electro-Chemie-Luminescence & TENG & Triboelectric Nanogenerators \\
		DIMG & Deep Inorganic Material Generator & GT4SD & General Toolkit for Scientific Discovery \\
		\hline
		\hline
	\end{tabular}\label{tab:abbriviation}
\end{table}


\section{Introduction}
Since the debut of graphene \citep{novoselov2004electric}, the landscape of material science has been forever altered, with the realm of two--dimensional (2D) materials emerging as an enduring focal point, captivating the research community for more than two decades now. Given its ubiquity and potential, nitrogen's prevalence alongside carbon renders CNs a prime candidate for cutting--edge research endeavors. Continuing its relevance to date, carbon nitrides owe much to their favorable experimental feasibility and diverse array of captivating attributes. From Dirac cone--like crossings to moderate semiconducting bandgaps, rich topological signatures to optical tunability across the whole visible spectrum, efficient thermal transport to surprisingly adept thermoelectric performances -- their multifaceted nature promises continued exploration and innovation \citep{zhao2015graphitic,wang2014topological,datta2020exploring,tan2021dirac,jana2023spontaneous,zhao2017exotic}. For example, the widely studied graphitic carbon nitride (g--C$_3$N$_4$) and its favorable attributes including its tunable optoelectronic properties has been applied in a diverse range of fields like photocatalysis, sensing and photovoltaics \citep{rono2021review,dong2016graphitic,liu2016graphitic}. The ongoing quest on this green energy material and its application has also been discussed in detail in different topical reviews \citep{miller2017carbon,wang2018recent,kong2021graphitic,chen2023graphitic,thomas2022synthesis}. Along with g--C$_3$N$_4$, structural intricacies and respective stoichiometry of other 2D CN materials and their functionalized versions have been extensively studied. In 2016 and 2017, the successive experimental large--scale synthesis of crystalline, hole--free, semiconducting single--layer of C$_3$N or famously known as 2D polyaniline by two different groups adapting distinct synthesis \citep{mahmood2016two,yang2017c3n} routes has notably pushed the limits further. It is a graphene--like honeycomb lattice with a homogeneous distribution of C and N atoms with a D$_6$h symmetry. Despite the vast majority of C$_x$N$_y$ materials, C$_3$N is the only one with a smaller indirect bandgap (0.39 eV) that exhibits ultrahigh stiffness (388 GPa--nm), low thermal conductivity (128 Wm$^{-1}$s$^{-1}$), low--temperature ferromagnetism and optical tunability throughout the visible spectrum with excitonic dark and bright states \citep{bonacci2022excitonic}. This equivalent nitride of graphene finds its distinctive ingenuity in multiple aspects -- from energy harvesting to nanoelectronics \citep{jiao2021surprisingly,bafekry2019c3n}. Monolayers like C$_2$N \citep{yong2019c2n,gao2022promoting}, C$_4$N \citep{li2024rectangular}, g--C$_4$N$_3$ \citep{tang2023exploring,he2023flat}, C$_6$N$_7$ \citep{zhang2024first,zhang2023first} are already established to be versatile and adaptable to various sectors whether it is academia or industry.

Apart from the single layer C$_x$N$_y$ networks, another very effective strategy to modulate and manipulate the material features has been the formation of heterojunctions or van--Der-Waals heterobilayers which facilitates easy exfoliation \citep{dong2015enhanced}. From an optoelectronic perspective, a very crucial factor is the presence of interlayer coupling which is necessary to mix different orbitals to justify the optical selection rules. For a g-C$_3$N$_4$ bilayer, it is reported that the primary bandgap gets increased by the interlayer coupling and the computed optical absorption depicts higher visible-light absorption \citep{wu2012visible}. In 2021, Wei \etal. \citep{wei2021bandgap} critically explored, both via theoretical and experimental techniques, the bandgaps of bilayers of two--dimensional C$_3$N which can be engineered by controlling the stacking order or applying an electric field. Synthesis of AA' and AB' stacked C$_3$N are found to have different bandgaps with controllable on/off ratios, high carrier mobilities and photoelectric detection abilities. Apart from that, an in-plane conjugated heterostructure of C$_2$N/C$_3$N or vertically stacked type--II C$_6$N$_6$/C$_3$N$_4$ with a reduced band gap turned out to be an efficient HER (hydrogen evolution reaction) \citep{xu2018design,liang2016photocatalytic}. A CN/C$_3$N$_2$ heterostructure has shown significant sensitivity towards SF$_6$ decomposed gases where one cell of this hetero--bilayer can capture eight of each gas with good adsorption strength \citep{yuan2024design}. Functionalizing the surface of the C$_x$N$_y$ networks by heteroatom doping or just a stable ground state of CNs distinctly morphed with different stoichiometries has shown rich features to ponder. The effect of heteroatom doping for modulating the electronic structure has enormously enhanced the catalytic and electrochemical activities of different carbon nitride systems and their distinctive hybrids \citep{fawaz2023emerging}. Along with others, Si doping has also been an efficient way to improve photocatalytic and electrocatalytic performance with suitable adsorption energies and activities \citep{roy2021graphitic,adekoya2021boosting,liu2015novel}. Emergence of a special `dumbbell' morphology is continuing to demand greater observation and recognition \citep{jana2023spontaneous,kaur2019energetics,duan2023diverse}. So this is really an ongoing quest.

Besides the conventional methods of study, the material science research is entering a brand new phase following the advent of artificial intelligence and machine learning (ML) \citep{morgan2020opportunities}. ML essentially uses huge amount of datasets to iteratively optimize models and make reasonable predictions. The burgeoning application of ML in semiconductor discovery and catalysis research may pave the way for green energy revolution \citep{kitchin2018machine}. Although the advancement of application of ML methods on CNs is in its initial stage, its prospect is undoubted \citep{yan2022development,mai2022machine}. Molecular dynamics simulations combined with ML algorithms are able to predict the free energy of exfoliation and solvation free energy for the liquid--phase exfoliation (LPE) of the g--C$_3$N$_4$ sheets \citep{shahini2023predicting}. It reduces the need to run additional lengthy computations. In fact, the identification of benzyl alcohol as a promising solvent was never incorporated previously for the LPE of 2D CNs. The bandgap of graphitic carbon nitride (GCN) subjected to external dopants under different experimental conditions are modeled using extreme machine learning (ELM) based models and hybrid support vector regression and genetic algorithm \citep{owolabi2021prediction} where all models utilize a specific surface area as a descriptor for estimating bandgap energy. This integration of ML methodologies with CNs has emerged as a key strategy to advance their applications across diverse fields of materials science. In catalysis, ML models play a crucial role in the rational design of CN based photocatalysts for applications such as environmental remediation and hydrogen generation, leveraging predictive capabilities to enhance efficiency and selectivity. Furthermore, ML-driven simulations facilitate the exploration of CNs in energy storage systems, optimizing parameters such as conductivity and charge storage capacities. In materials discovery, ML algorithms aid in the systematic exploration and discovery of novel CN structures with tailored properties, thereby accelerating the innovation cycle in material science. This synergistic approach underscores the transformative potential of integrating ML with CNs, paving the way for advancements in sustainable technology and materials development.

In this topical review, we have aimed to critically discuss the arrival of 2D carbon nitrides, their ever--increasing popularity, high synthetic feasibility as well as the application perspectives alongside their future prospects focusing on the highly manifesting power of ML modelling. We will focus on how this technique can help to speed--up, predict and compute different aspects of this family of materials that can unquestionably push the boundaries further. Table-\ref{tab:abbriviation} displays all the abbreviations and subsequent definitions that we have used throughout the article.
\section{All about carbon nitrides}
\begin{figure*}[h!]
    \centering
    \includegraphics[scale=0.485]{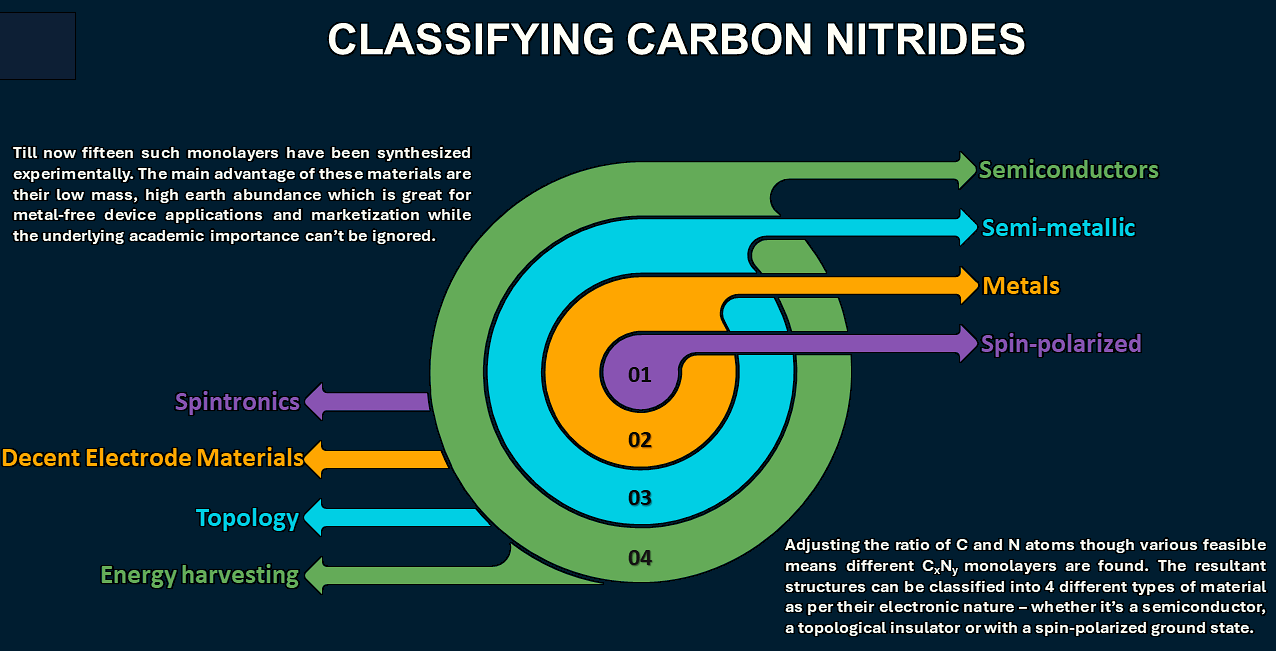}
    \caption{A schematic classification of existing monolayer carbon nitrides (both experimentally synthesized and theoretically proposed) based on their distinct electronic nature from semiconducting, metallic to semi-metallic and magnetic (or spin-polarized) ground states with plausible genres of application.} 
    \label{Schematic}
\end{figure*}

Now delving into the details of different CNs whether 2D monolayers or distinctly stacked bilayers, we will classify all existing C$_x$N$_y$ materials as per their electronic nature. Tuning the C$:$N proportion, introducing structural defects, through heteroatom doping, different lattice symmetries or stacking monolayers consequently lead to rich electronic structure - from metals and wide bandgaps to half-metallicity or topologically protected semi-metals as shown in Figure-\ref{Schematic}.
\subsection{\textbf{Semiconductors}}
A single layer of CN with a 1$:$1 proportion of C and N, namely graphitic nanoporous CN has been synthesized successively via the reactions of C$_3$N$_3$Cl$_3$ and Na by a simple solvothermal method \citep{li2006self} or by a simple polymerization of 3-amino-1,2,4-triazine \citep{mane2013selective}. As far as the lattice symmetry and parameters are concerned, single layer and non-magnetic g-CN possesses a hexagonal unit cell with equal a and b where a$=$b$=$7.118 \AA{} with an underlying symmetry group of P6/mmm. The g-CN monolayer has a direct band gap of 1.57 eV (PBE) and 3.18 eV (HSE06) \citep{srinivasu2014porous} whereas the experimentally realized bandgap shows a value of 2.73 eV \citep{li2006self,mane2013selective}. It appears to have a very high surface-to-volume ratio and decent physiochemical stability \citep{wang20212d,kumar2021single}, as it has broad application prospects in the field of hydrogen storage \citep{chen2021computational} and electrocatalyst \citep{chu2021single}. There is also a sp$^3$ hybridized three-sublayer crystal structure with four-coordinated carbon atoms, namely Tetra-Hexa CN$_2$ sandwiched between two sublayers of three-coordinated nitrogen atoms. The rectangular unit cell contains 12 atoms - four C and eight N atoms. The optimized lattice constants are a = 4.18 Å and b = 5.78 Å. The buckling length between the two N layers is 1.48 Å. This material is isostructural to the tetra-hexa-carbon \citep{ram2018tetrahexcarbon} but with reduced lattice constants and increased buckling. Electronically, TH-CN$_2$ is a wide indirect gap semiconductor with a value of 4.57 eV as per the more accurate HSE06 functional \citep{wei2021new}. A similar stoichiometric monolayer, Penta CN$_2$ which is a nitrogen-rich (nearly 21.66\%) material with a tetragonal lattice with a lattice constant a = 3.31 Å and belongs to a crystal group of P-42$_1$m layer group number-58 \citep{zhang2016beyond}. The relaxed structure of penta-CN$_2$ is composed of three layers of C atoms in the middle and N atoms on the top and bottom \citep{zhang2015penta} - analogous to well-known B$_2$C (1.08 \AA{}) \citep{li2015flexible} or SiC$_2$ (1.33 \AA{}) \citep{lopez2015sigma} etc. This single layer also possesses a large indirect bandgap semiconductor with a bandgap value of 4.83 eV (PBE) and 6.53 eV (HSE06). Penta-CN$_2$ also displays an interesting double degeneracy along the first Brillouin zone edges, which is topologically protected by the nonsymmorphic symmetry of the structure.

The hexagonal unit cell of C$_2$N is constructed in chain-like rings and the hexagonal rings are linked via the C−N bond. The optimized lattice constant of the C$_2$N monolayer is estimated to be of 8.33 \AA{}. In 2015, Mahmood \etal reported the preparation of layered two-dimensional C$_2$N via a simple wet-chemical reaction \citep{mahmood2015nitrogenated}. It was prepared by mixing hexa-amino-benzene (HAB) trihydrochloride and hexa-keto-cyclohexane (HKH) octahydrate in N-methyl-2-pyrrolidone. The consequent material was dark black and exhibited experimental direct bandgaps of 1.96 eV which closely agrees with the theoretical calculations (1.7 eV). Figure-\ref{C2N} below displays all the necessary experimental findings. 
\begin{figure*}[h!]
    \centering
    \includegraphics[width=16cm]{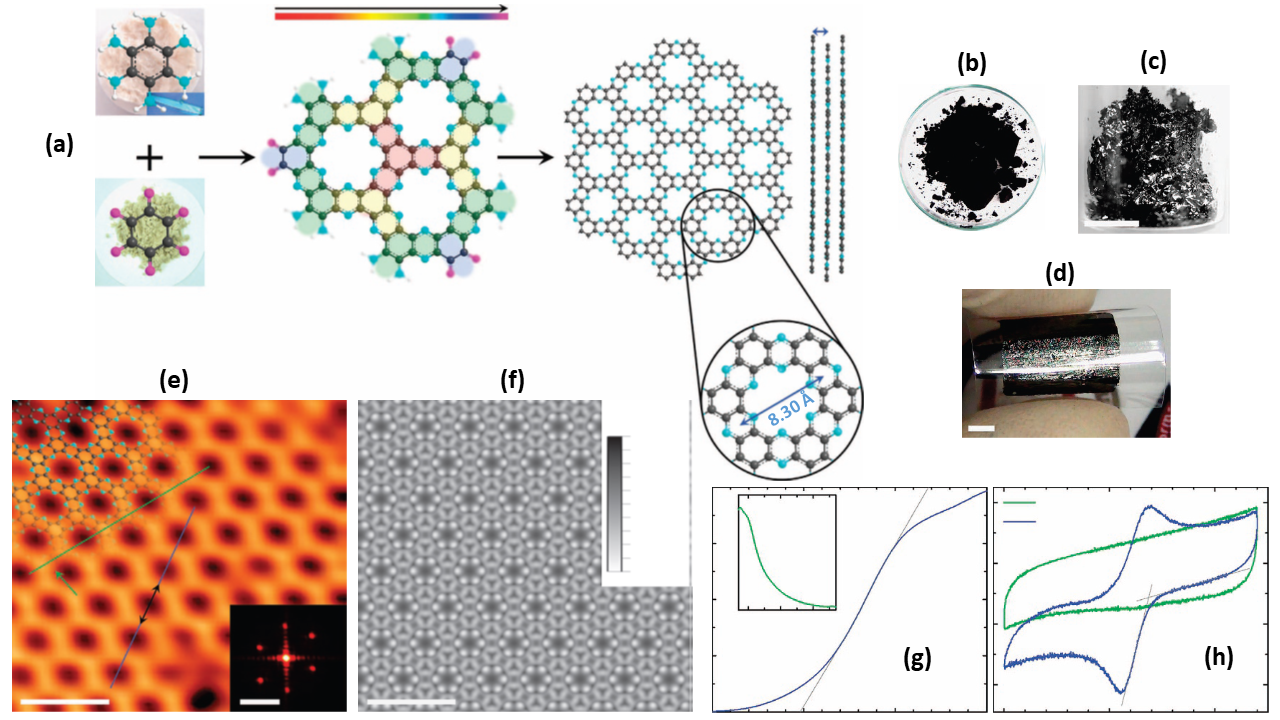}
    \caption{(a) Schematic representation of the reaction between hexaaminobenzene (HAB) trihydrochloride and
hexaketocyclohexane (HKH) octahydrate to produce the C$_2$N-h2D crystal. The inset in the image of HAB is a polarized optical microscopy image of the
HAB single crystal. Digital photographs: (b) as-prepared C$_2$N-h2D crystal; (c) solution-cast C$_2$N-h2D crystal on a SiO$_2$ surface after heat-treatment at
700$^o$C; (d) a C$_2$N-h2D crystal film (thickness: approximately 330 nm) transferred onto a PET substrate. The shiny metallic reflection of the sample
indicates that it is highly crystalline. Figure-(e,f) shows the atomic-resolution STM topography image of the C$_2$N-h2D crystal on Cu(111). The STM image was obtained
at a sample bias of 0.7 V and a tunelling current of 300 pA. The top-left inset is the structure of the C$_2$N-h2D crystal superimposed on the image.
The bottom-right inset is 2D fast Fourier transform and the simulated image respectively. Figure-(g,h) show the results of optical band-gap measurements and a plot of the absorbance squared vs.
photon energy extrapolated to zero absorption. The inset is the ultraviolet absorption curve. (b) Cyclic voltammograms of the C$_2$N-h2D crystal at a scan rate of 100mVs$^{-1}$ using a Ag/Ag$^{+}$ reference electrode. Reprinted with permission from \citep{mahmood2015nitrogenated} Copyright 2015, Nature}
    \label{C2N}
\end{figure*}

Ever since its successful synthesis, reports on using C$_2$N materials have flourished quite drastically \citep{tian2020c2n}. Lei Wang \etal reported a
metal-free Z-scheme preparation for water splitting using C$_2$N and aza-fused microporous polymeric nanosheets in 2018 \citep{wang2018van}, highlighting the potential of the polymer photocatalyst. This Z-scheme plays a vital role in photocatalytic reactions that require both reducing and oxidizing capabilities. For instance, in photocatalytic water splitting, light energy drives the separation of water into hydrogen and oxygen. The Z-scheme facilitates efficient separation and transfer of photogenerated charge carriers between two different semiconductors, enabling the simultaneous reduction of protons to hydrogen and oxidation of water to oxygen. This mechanism is crucial for designing efficient photocatalytic systems for applications such as solar fuel production and environmental remediation.  Especially, C$_2$N has been reported to be quite superior in terms of elasticity compared to the other two allotropes, namely heptazine and s-triazine \citep{abdullahi2018elastic}. A recent report has displayed its tetragonal counterpart namely T-C$_2$N which has a relaxed lattice parameter of 5.99 \AA{} and electronically possesses a smaller indirect bandgap of about 0.73 eV \citep{xue2023novel}. The unstable puckered N nanosheet (i.e. bp-N monolayer) has been stabilized by inserting C$_2$ dimers between the upper and lower N zigzag lines which in turn forms C$_2$N$_2$ nanosheet possesses good energetic, dynamical and thermal stability \citep{wang2020stable}. The relaxed unit cell of C$_2$N$_2$ sheet is orthogonal having lattice parameters of 3.46 and 2.39 \AA{} in two different directions with a space group of Pmna (53). This is an anisotropic indirect gap semiconductor of value 2.56 eV (SCAN) and 3.58 eV (HSE06) with simultaneous traces of ultrahigh carrier mobilities and large in-plane anisotropy. A large anisotropic mobility ratio of 85 for the hole mobility of bp-C$_2$N$_2$ nanosheet is discovered - the maximum anisotropy value reported for a 2D material.

A C$_2$N$_3$ unit cell has an optimized lattice parameter of 7.864 \AA{} which has a total 12 C atoms and 18 N atoms with the underlying P6/mcm space group \citep{shi2018mechanical}. It shows a heavily doped p-type semiconductor characteristic with an indirect bandgap of 4.62 eV (in HSE06) and 3.26 eV as per PBE functional. In 2017, Miller \etal synthesized single-layer of CN nanosheets with an approximately close stoichiometry of C$_2$N$_3$ via the spontaneous dissolution of bulk PTI based carbon nitride \citep{miller2017single}.
 
\begin{figure*}[h!]
    \centering
    \includegraphics[scale=0.48]{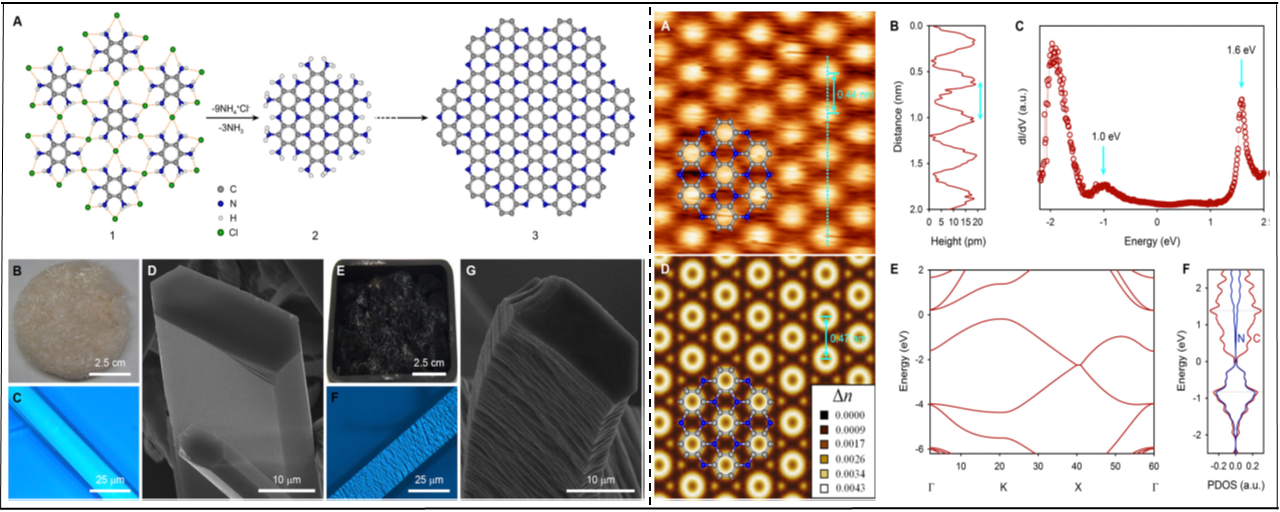}
    \caption{In the left panel the schematic representation of 2D PANI formation is shown here. (A) Single-crystal X-ray packing structure of HAB (structure-1); structure of 2D PANI unit with edge groups (C$_3$NH, structure-2), and the spontaneous transformation of HAB crystal unit into the 2D PANI structure (structure-3). Morphology changes of HAB crystals into 2D PANI frameworks. (B) Digital photograph of HAB crystals on butter paper. (C) Optical microscopy image of a needle-like HAB crystal before annealing. (D) SEM image of an HAB single crystal before annealing. (E) Digital image of HAB after annealing at 500°C. (F) Optical microscopy image of 2D PANI crystal after annealing at 500°C. (G) SEM image of a 2D PANI single crystal after annealing. STM and theoretical studies of the 2D PANI structure. In the right panel - (A) STM image of a 2D PANI framework (2.5 × 2.5 nm $^2$, V$_s$ = −1.1 V, I$_t$ = 1.0 nA). Inset structure represents C$_3$N repeating unit with carbon atom (gray ball) and nitrogen atom (blue ball). (B) Topographic height profile along the cyan dot line marked in A. (C) Differential conductance (dI/dV) spectrum of a 2D PANI framework. (D) Simulated STM image with superimposed structure of C$_3$N repeating unit. (E) Electronic band structure. (F) PDOS of the carbon (dark red) and nitrogen (dark blue) atoms. Reprinted with permission from \citep{mahmood2016two} Copyright 2016 PNAS.
} 
    \label{C3N}
\end{figure*}

For the family C$_3$N$_x$ (x = 1-7), Baek and co-workers reported the synthesis of 2D C$_3$N from the direct pyrolysis of hexa-amino-benzane trihydrochloride (HAB) single crystals at 500°C for 2 hr \citep{mahmood2016two} as shown in Figure-\ref{C3N} with topographical images and underlying electronic spectra. They studied the C$_3$N formation mechanism in detail, which involved the evolution of ammonia and ammonium chloride. Yang \etal reported a controllable large-scale synthesis of C$_3$N quantum dots by 2,3-diaminophenazine polymerization through hydrothermal process \citep{yang2017c3n}. The polymerization steps were identified by Matrix-Assisted Laser Desorption Ionization - Time of Flight Mass Spectroscopy (MALDI-TOFSMS). The well-crystallized unit cell of graphene-like C$_3$N possesses a honeycomb structure with a lattice constant of 4.86 \AA{} and the two-fold hexagonal single crystal form indicates D$_6$h-symmetry of the N and C atoms. It is a small bandgap indirect semiconductor (1.23 eV in HSE06 and 0.39 eV in PBE) with ultrahigh stiffness (higher than graphene) and traces of low-temperature ferromagnetism (below 96 K) with hydrogen doping. 

Its tetragonal counterpart possesses a wide and direct semiconducting gap of 5.74 eV with a proper pore size of 5.5 \AA{} - perfect for water infiltration \citep{zhou2020novel}. In a recent work of ours, we have critically analyzed and reported spontaneous dumbbell formations from monolayer C$_3$N which contains three indirect gap semiconductors (about 2-3 eV) of dumbbell C$_3$NX (X = B, P and As) \citep{jana2023spontaneous}.

Porous graphitic C$_x$N$_y$ monolayers with semiconducting features have drawn wider attention over the years because of built-in pores with active sites and larger surface area. Only a few such CNs have been synthesized - covering the bandgap range of 1.23–3.18 eV. A recent study systematically investigates two new 2D monolayers Tetra- and hexa-C$_3$N$_2$ with realistic pathways for their experimental realization \citep{cai2023c}. The relaxed structures of H- and T-C$_3$N$_2$ comprise lattice parameters of 12.29 \AA{} and 8.63 \AA{} with subsequent space groups P6/mmm and P2/m respectively. The underlying electronic depicts the direct bandgap nature of about 0.35 eV (T-C$_3$N$_2$) and 1.97 eV (H-C$_3$N$_2$) as per HSE06 functional which are nearly double that of in PBE with high carrier mobilities and excellent visible-light absorption.

For the graphitic C$_3$N$_3$ cell, every six C$_3$N$_3$ rings enclose a pore in C$_3$N$_3$ layer. The C-C and C-N bond lengths are 1.51 \AA{} and 1.34 \AA{} respectively with the optimized lattice parameters of 7.11 \AA{}. The diameter of the pore is 5.46 \AA{}, almost the same as that of the one in graphdiyne (5.42 \AA{}). The pore edge is surrounded by 6 N atoms with zero dangling bonds. The band structure shows a direct band gap of 1.60 eV. This assembled CN network has also been realized experimentally through a simple solvothermal technique with well-controlled dimensionality and luminescence \citep{li2006self}. Carbon nitride nanotube bundles were formed using NiCl$_2$ as a catalyst precursor while the reaction of cyanuric chloride (C$_3$N$_3$Cl$_3$) with sodium at 230°C and 1.8 MPa is taking place in a stainless steel autoclave.

\begin{figure*}[h!]
    \centering
    \includegraphics[width=16cm]{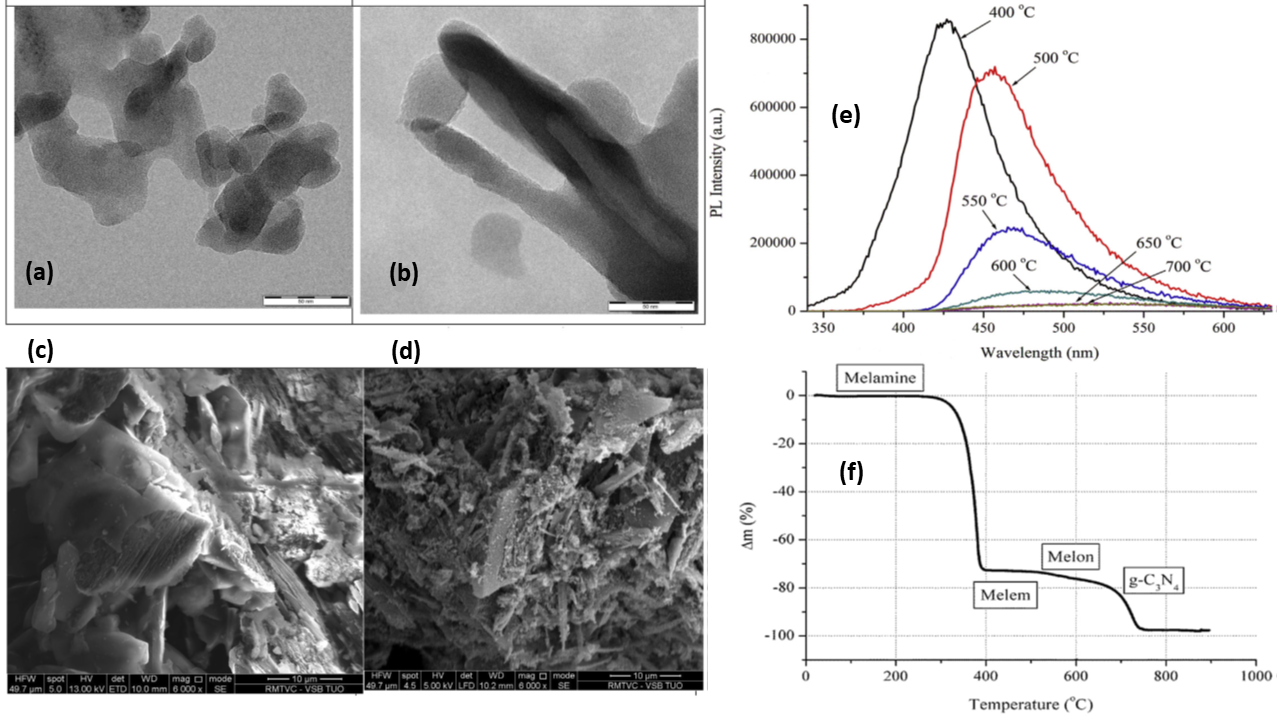}
    \caption{Figure-(a,b) and (c,d) shows the TEM and SEM images of g-C$_3$N$_4$ samples obtained at 600$^o$C and 700$^o$C. The photoluminescence spectra of melamine products obtained at different temperatures and the thermogravimetric curve of annealing of melamine are shown in figure-(e) and (f) respectively. Reprinted with permission from \citep{praus2017graphitic} Copyright 2017 Elsevier} 
    \label{C3N4_mel}
\end{figure*}

\begin{figure*}[h!]
	\centering
	\includegraphics[width=15.5cm,height=15cm]{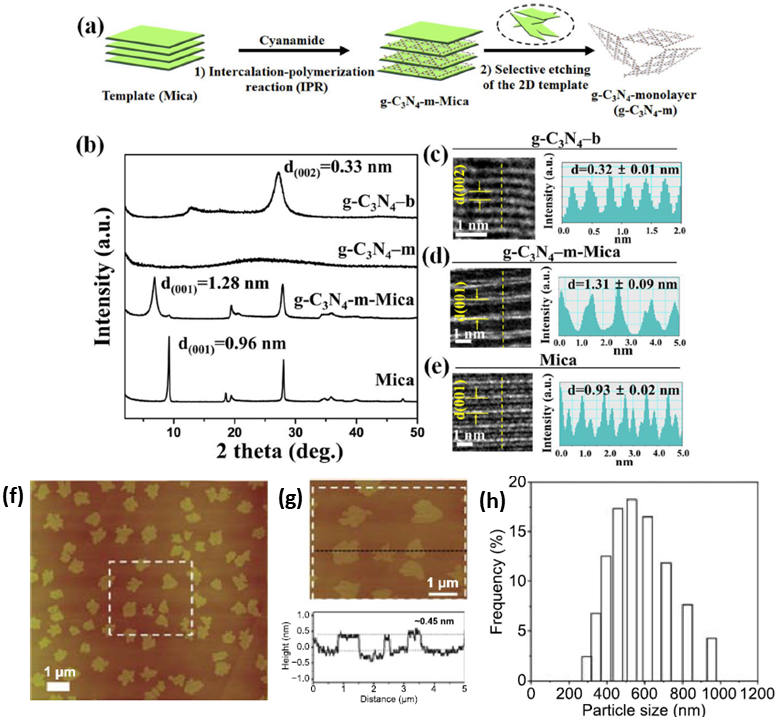}
	\caption{(a) Schematic diagram for the synthesis of g-C$_3$N$_4$–m by using the 2D Mica template. (b) XRD patterns of Mica, g-C$_3$N$_4$–m–Mica, g-C$_3$N$_4$–m and g-C$_3$N$_4$–b. (c–e) Cross-sectional HRTEM images and photometric intensity profiles along the yellow dashed lines for g-C$_3$N$_4$–b, g-C$_3$N$_4$–m–Mica and Mica. \textbf{AFM image of g-C$_3$N$_4$–m}.  (f)The g-C$_3$N$_4$–m monolayers dispersed on the freshly-cleaved muscovite mica substrate. (g) The thickness of g-C$_3$N$_4$–m monolayer flakes, and the height profile along the black dashed line. (h) The histogram for the size distribution of g-C$_3$N$_4$–m. Reproduced with permission from Springer Nature (2022).}
	\label{2D-C3N4}
\end{figure*}

Graphitic carbonic nitride (g-C$_3$N$_4$) has attracted much attention since it was first constructed to be a visible-light-driven photocatalyst due to its high abundance, stability and decent capacity for solar utilization \citep{wang2009metal}. In the realm of energy conversion and storage, the unique layered structure of g-C$_3$N$_4$, along with its tunable bandgap, metal-free nature, high physicochemical stability, and ease of synthesis, have significantly contributed to its growing popularity including batteries and supercapacitors. \citep{luo2019graphitic,ghosh20222d,wang20212d}. The porous structure plays a crucial role in enhancing the performance of negative electrodes in lithium-ion batteries. Pan \etal explored g-C$_3$N$_4$ nanotubes with porous architectures through a first-principles study. These porous structures shorten the diffusion pathways for Li+ ions during lithiation and delithiation processes, enabling the material to achieve a relatively high specific capacity of up to 1165.3 mAh/g \citep{pan2014graphitic}.. However, its high nitrogen content has led to challenges such as poor conductivity and substantial irreversible capacity loss. To address these issues, Chen \etal introduced a magnesiothermic denitriding technique to reduce the nitrogen content of g-C$_3$N$_4$, enhancing its performance as a lithium-ion battery anode \citep{chen2017nitrogen}. This approach achieved a highly reversible lithium storage capacity of 2753 mAh/g after 300 cycles, with improved cycling stability and rate capability. Several phases of C$_3$N$_4$ ($\alpha, \beta, \gamma$ or cubic) with their structural formation, semiconducting electronic nature and photocatalytic activities have been reported and studied earlier \citep{datta2020exploring}. A facile one-pot synthesis of nanoporous graphitic carbon nitride has been done using different soft and direct templates through the self-polymerization reaction of dicyandiamide (DCDA) \citep{wang2010facile}. This material was also realized experimentally by condensing melamine in air for 2 hr and the exfoliation is observed beyond 600°C. It was used for the photocatalytic decomposition of N$_2$O \citep{praus2017graphitic}. Figure-\ref{C3N4_mel} depicts the scanning and tunelling electron microscope (SEM and TEM) images for g-C$_3$N$_4$ samples obtained at 600°C and 700°C. In 2022, g-C$_3$N$_4$ monolayer in the perfect 2D limit was successfully realized, for the first time, by the well-defined chemical strategy based
on the bottom-up process \citep{piao2022monolayer}. In Figure-\ref{2D-C3N4} we can see all the details which includes schematic diagram of the synthesis using 2D Mica plate, XRD patterns, HRTEM, AFM imaging with the subsequent histogram for the size distribution of g-C$_3$N$_4$-m.

Inspired by the polymerization of urea, the preparation of the novel C−C bridged heptazine CNs UOx (x is the ratio of urea to oxamide, x = 1, 1.5, 2, 2.5 and 3) has been achieved, which is close to (C$_6$N$_7$)$_n$ for the first time by introducing part of oxamide. A simple and efficient process
for the synthesis of 2D CNs and related materials of very high quality is reported which relies on the use of a metallic surface as both a reagent and a support for the coupling of small halogenated building blocks \cite{moreira2023simple}.

Now before delving into further results and their subsequent implications, the connection between porosity density and the ratio of nitrogen atoms in carbon nitride (CN) lattices needs to be addressed as it is quite crucial for understanding and optimizing the properties of these materials. The porosity density directly influences the available surface area and the pathways for ion diffusion, which are important for applications in energy storage \citep{gong2015carbon}, catalysis \citep{han2023nitrogen}, and sensing \citep{patel2023mesoporous}. A higher porosity typically enhances the material’s capacity to adsorb and interact with molecules, improving its performance in catalysis and sensor applications. The ratio of nitrogen atoms in CN lattices also plays a significant role in modulating electronic properties \citep{wang2022synthesis}. Nitrogen doping in carbon materials can create localized states in the electronic structure, which can affect conductivity, charge storage capabilities, and catalytic activity. By adjusting the nitrogen content, it may be possible to tune the electronic structure, making CN materials more suitable for specific applications. For instance, higher nitrogen content can enhance the electrochemical performance of materials used in energy storage, as nitrogen can help stabilize lithium or sodium ions in batteries \citep{kumar2024doping,shi2020nitrogen}. Even from a mechanical standpoint, the porosity and nitrogen content can affect the material’s stability, strength, and elasticity. In porous CN materials, the mechanical properties are often influenced by the distribution and connectivity of nitrogen atoms within the lattice, as well as by the pore size and structure. Nitrogen-rich graphitic carbon nitrides (g-C$_3$N$_5$) have emerged as promising star for photocatalytic applications due to its significant enhancements in light absorption properties, which can activate in ultraviolet, visible, and even under near-infrared irradiation \citep{vuong2023nitrogen}. This can also enhance the material's resistance to thermal degradation, making it more stable under high-temperature conditions, which is beneficial for catalysis, hydrogen evolution and energy storage applications \citep{wang2020new,skorupska2021effect,khosravifar2024nitrogen}. In 2021, Liu \etal reported a series of 2D co-catalyst nitrogen-doped graphenes with the different bond states of doped nitrogen were synthesized under modified synthetic conditions, resulting in different surface catalytic hydrogen peroxide generation activities \citep{liu2021nitrogen}. Table-2 properly showcases all the structural details, underlying symmetry, subsequent electronic nature and synthesis (if yet synthesized or not) route taken with proper referencing.

\subsection{\textbf{Semi-metallics:}}The emergence of Dirac points in 2D materials, or more specifically, in 2D CNs which are typically insulating, is rare. Moreover, the appearance of symmetry-protected nodal-line semimetal with sheer robustness against external strain is quite surprising. For example, in 2020, Tan \etal discovered a family of holey nitride monolayers namely C$_x$N$_3$ (x= 7,10,13,19) with Dirac-like crossing exactly at the Fermi level. It has been constructed from a $\sqrt{13}\times\sqrt{13}$ graphene by removing a six-membered ring of carbon atoms - replacing the carbon atoms at the edge of the hole with nitrogen atoms. These four novel nanosheets belong to two distinct groups, namely space group P6mm (183) for C$_{10}$N$_3$ and C$_{13}$N$_3$ and P6 (168) for the other two structures. Another interesting occurrence is the traces of intriguing topological signatures in low-mass structural forms due to their weaker spin-orbit interaction. In recent years, a pool of such materials has been discovered with rich electronic structures carrying symmetry-protected semi-metallicity or quasi-1D Dirac nodal lines and topological flat edge states such as zigzag buckled C$_{4}$N \citep{li2024rectangular}, dumbbell C$_3$NX (X = C,Si,Ge) \citep{mondal2024intriguing}, C$_x$N$_4$ (x = 4,5,9) \citep{li2019novel,li2022novel,chen2018prediction}, C$_{19}$N$_3$ \citep{li2015new} etc. It has been suggested that experimentally the 1,3,5,7-Tetrazocine monomer and the dehydrogenation reaction of C$_2$H$_4$ and C$_4$H$_4$N$_4$ can be used to synthesize semimetallic C$_4$N$_4$ and C$_5$N$_4$ respectively. A very recent article reported a strong anisotropic Dirac state in monolayer TPH-C$_5$N$_3$ with non-zero $\mathbb{Z}_2$ invariant and topologically non-trivial edge states \citep{tan2024strong}. A possible synthetic pathway has also been proposed through dehydrogenation and polymerization while cyclobutadiene, pyrazine and tetra-aminoethylene are chosen as precursors. Table-2 has all the necessary details and subsequent references.

\subsection{\textbf{Metals:}} One of the most recent members of the carbon-nitride family, monolayer C$_5$N, reported to be chemically, mechanically and thermodynamically stable is found to be metallic and reported to be a promising building block for the anode of Potassium-ion batteries \citep{jin2022monolayer}. In 2011, 2D C$_5$N$_2$ was synthesized successfully by Kou \etal \citep{kou2011supercapacitive} with a large surface area and high conductivity. Synthesis of such porous frameworks based on aza-fused CMPs were ionothermally realized by condensation of 1,2,4,5-benzenetetramine with triquinoyl hydrate from 300 to 500$^0$C. In this year, allotropes of three novel C$_5$N$_2$ (type-a, b and c) are shown to have great potential as high-performance alkali metal ion battery material \citep{you20242d}. Recently, a planar and hybridized metallic structure with h57 Haeckelite orientation C$_7$N is proposed with consequent application in high-capacity anode for post-Li-ion batteries \citep{hajiahmadi2023first}. Apart from that, for C$_9$N$_4$ and C$_{10}$N$_{3}$, the Fermi level is found to be crossed by the $\sigma$ (s+p$_x$+p$_y$) and $\pi$ (p$_z$) orbitals of N and C atoms, leading to the formation of metallic character \citep{mortazavi2019prediction}.

\subsection{\textbf{Spin-polarized:}} Half-metals - displaying the feature of a metal in one spin channel and of a semiconductor in the other, have attracted some serious research attention due to their potential applications in the spintronic field over the years \citep{wolf2001spintronics,he2010robust}. As far as the low-dimensional CNs are concerned, very few have been reported over the course of time. In the year 2010, a graphitic carbon nitride, namely, g-C$_4$N$_3$ was experimentally realized by Lee \etal \citep{lee2010fluidic} which interestingly can exhibit half-metallicity without any external modification and preserves the formational and magnetic stability even at 500 K \citep{du2012first}. Functional porous carbons were prepared via direct, ambient-pressure, thermal pyrolysis of task-specific ionic liquids (ILs). This process lies in the synergistic usage of the insignificant volatility of the ILs and the inclusion of the cross linkable nitrile groups in the anions. The resulting product retained an extremely high N content of about 18\% at a temperature of 800$^0$C. Incorporating hydrogen dangling bonds has also been reported to be a decent strategy to tune the magnetic features of metal-free g-C$_3$N$_4$ nanosheets \citep{xu2015hydrogen}. A unique form of monolayer viz. dumbbell-C$_7$N$_3$ can be formed after substituting three N atoms in the dumbbell-C structure which showcases magnetism and the magnetic moment bears a value of 1 $\mu$B. This is a specifically Bipolar magnetic semiconductor (BMS), where the valence and conduction bands possess opposite spin polarization with different bandgap values (direct gap of 0.9 eV for up-spin and indirect of 1.7 eV for spin-down). This type of material can control the gate voltage properly. Gao \etal in 2020 showed the emergence of ferromagnetism in graphitic carbon nitrides through nitrogen defects \citep{gao2020ferromagnetism}. Table-1 here shows all different existing magnetic graphitic C$_x$N$_y$ with necessary structural details with corresponding electronic nature whether a half-metal (HM), BMS or spin-polarized metal (SPM).
\begin{table*}[h!]
\centering
\label{structure}
\caption{Table containing the basic details of different carbon nitride networks. Here SC, M, SPM, SM, DNL, HM, BMS stand for the semiconducting, metallic, spin-polarized metal, semi-metallic, Dirac nodal line, half-metallic and bipolar magnetic semiconducting electronic nature of the materials respectively.}
\resizebox{\textwidth}{!}{\begin{tabular}{c c c c c c c c c c c c c}
\hline
\hline
Structure & a$_0$ (in \AA{}) & Symmetry & Buckling (in \AA{})& Electronic nature (Type)& Bandgap (in eV - Theory/Expt.)& Synthesis (Method)&Reference\\
\hline
\hline
CN&7.12&Hexagonal&-&SC (Direct)&3.18/2.73&Y (Solvothermal or Polymerization)&\citep{li2006self,yong2022highly,mane2013selective}\\
Tetra-Hexa CN$_2$&4.18 (and 5.78)&Rectangular& 1.48 & SC (Indirect)&4.57/-&-&\citep{wei2021new}\\
Penta-CN$_2$&3.312&Tetragonal&1.52&SC(Indirect)&6.53/-&-&\citep{zhang2016beyond,wang2022penta}\\
C$_2$N&8.33&Hexagonal&-&SC(Direct)&1.7/1.96&Y(Wet-chemical reaction)&\citep{mahmood2015nitrogenated,tian2020c2n}\\
T-C$_2$N&5.99&Tetragonal&-&SC(Indirect)&0.73/-&-&\citep{xue2023novel}\\
C$_2$N$_2$&3.46 (and 2.39)&Orthogonal&3.10&SC(Indirect)&3.58/-&-&\citep{wang2020stable}\\
C$_2$N$_3$&7.864&Hexagonal&-&SC(Indirect)&4.62/-&Y (Spontaneous dissolution)&\citep{shi2018mechanical,miller2017single}\\
C$_3$N&4.86&Hexagonal&-&SC(Indirect)&1.23/0.39&Y(Direct pyrolysis)&\citep{wei2021bandgap,mahmood2016two,yang2017c3n}\\
T-C$_3$N&3.38&Tetragonal&-&SC(Direct)&5.74/-&-&\citep{zhou2020novel}\\
Dumbbell C$_3$NX&4.76-4.88&Hexagonal&1.92-3.62& SC (Indirect) and SM&2-3&-&\citep{jana2023spontaneous}\\
H-C$_3$N$_2$&12.29&Hexagonal&-&SC (Indirect)&1.97/-&-&\citep{cai2023c}\\
T-C$_3$N$_2$&8.63&Hexagonal&-&SC (Direct)&0.35/-&-&\citep{cai2023c}\\
g-C$_3$N$_3$&7.11&Hexagonal&-&SC (Direct)&1.6/2.2&Y (Solvothermal)&\citep{guo2005synthesis,li2006self,ma2014computational}\\
g-C$_3$N$_4$&4.79&Hexagonal&-&SC (Indirect)&2.76/2.7&Y (Condensation/ Template)&\citep{praus2017graphitic,wang2010facile,datta2020exploring}\\
g-C$_3$N$_5$&15.203&Hexagonal&-&SC (Direct)&2.19/1.76&Y(Thermal deammoniation)&\citep{singh2022strain,liu2020preparation,kumar2019c3n5}\\
s-C$_3$N$_6$&10.57&Hexagonal&-&SC(Direct)&2.59/-&Y (Low temperature pyrolysis-Nanotemplating)&\citep{mortazavi2020nanoporous,kim2020thermodynamically}\\
C$_3$N$_7$&7.69 (and 6.37)&Triclinic&-&SC (Direct)&0.82/-&Y (Low temperature pyrolysis-Nanotemplating)&\citep{kim2020thermodynamically}\\
C$_4$N&12.64&Hexagonal&-&SC (Direct)&1.80/2.55&Y (Hydrothermal solvent)&\citep{yang2019theory}\\
C$_4$N-I&5.44&Rectangular&-&SC (Direct)&0.09/-&-&\citep{pu2017two}\\
C$_4$N-II&4.45&Rectangular&-&SM&-/-&-&\citep{pu2017two}\\
ZB C$_4$N-I&4.747 (and 4.114)&Rectangular&1.55&DNL&-/-&-&\citep{li2024rectangular}\\
ZB C$_4$N-II&4.741 (and 4.114)&Rectangular&1.54&DNL&-/-&-&\citep{li2024rectangular}\\
g-C$_4$N$_3$&4.81&Hexagonal&-&HM(Direct)&2.0/2.0&Y (Thermal pyrolysis)&\citep{lee2010fluidic,du2012first}\\
C$_4$N$_4$&3.58 (and 6.10)&Rectangular&0.41&SC (Direct)&0.68/-&-&\citep{li2019novel}\\
C$_4$N$_4$&4.99&Square&-&SM&-/-&-&\citep{huang2023c4n4}\\
C$_5$N&3.65 (and 8.90)&Rectangular&-&M&-/-&-&\citep{jin2022monolayer}\\
C$_5$N$_2$ (type - a, b and c)&6.41&Hexagonal&-&M&-/-&-&\citep{you20242d}\\
TPH-C$_5$N$_3$&7.79 (and 6.05)&Rectangular&-&SM&-/-&-&\citep{tan2024strong}\\
C$_5$N$_4$&7.19&Orthorhombic&-&SM&-/-&-&\citep{li2022novel}\\
C$_6$N&7.19 (and 8.01)&Rectangular&1.71&SM&-/-&-&\citep{bafekry2021two}\\
C$_6$N$_2$&7.31 (and 5.76)&Rectangular&-&SM&-/-&-&\citep{he2020novel}\\
C$_6$N$_3$&11.07&Hexagonal&-&M&-/-&-&\citep{yuan2022c6n3}\\
Dumbbell C$_6$N$_4$&4.64&Hexagonal&2.04&SC (Direct)&0.80/-&-&\citep{duan2023diverse}\\
C$_6$N$_6$&7.12&Hexagonal&-&SC (Direct)&3.22/-&-&\citep{srinivasu2014porous,li2019charge,wang2014topological}\\
C$_6$N$_7$&11.72&Hexagonal&-&SC (Direct)&1.97/2.09&Y (Thermal polymerisation)&\citep{zhao2021new,bafekry2022tunable}\\
C$_7$N&10.0 (and 5.8)&Rectangular&-&M&-/-&-&\citep{hajiahmadi2023first}\\
Dumbbell C$_7$N$_2$&4.89&Hexagonal&1.67&M&-/-&-&\citep{duan2023diverse}\\
Dumbbell C$_7$N$_3$&4.89&Hexagonal&2.02&BMS (Direct for $\uparrow$ and indirect for $\downarrow$)&0.90($\uparrow$) and 1.70($\downarrow$)/-&-&\citep{duan2023diverse}\\
C$_7$N$_3$&8.69&Hexagonal&-&SM&-/-&-&\citep{tan2021dirac}\\
C$_7$N$_6$&6.79&Hexagonal&-&SC (Direct)&2.25/-&-&\citep{hu2021c7n6}\\
C$_7$N$_9$&7.19&Hexagonal&-&SPM&-/0.006&-&\citep{gao2020ferromagnetism}\\
g-C$_8$N$_6$&7.17&Hexagonal&-&SC (Indirect)&2.89/-&-&\citep{li2020prediction}\\
C$_8$N$_8$&7.23&Tetragonal&-&M&-/-&-&\citep{kokabi2023noble}\\
Dumbbell C$_9$N&4.845&Hexagonal&1.72&M&-/-&-&\citep{duan2023diverse}\\
C$_9$N$_4$&6.875&Hexagonal&-&M&-/-&-&\citep{mortazavi2019prediction}\\
C$_9$N$_4$&9.64&Hexagonal&-&SM&-/-&-&\citep{chen2018prediction}\\
C$_9$N$_7$&7.23&Hexagonal&-&HM (Direct)&0.29/0.46&-&\citep{gao2020ferromagnetism}\\
C$_{10}$N$_{3}$&6.948&Hexagonal&-&M&-/-&-&\citep{mortazavi2019prediction}\\
C$_{10}$N$_{6}$&7.29&Hexagonal&-&HM (Direct)&1.52/2.72&-&\citep{gao2020ferromagnetism}\\
C$_{12}$N&-&Hexagonal&-&SC (Direct)&0.62/0.98&- &\citep{xiang2012ordered}\\
C$_{12}$N$_{2}$&7.985&Hexagonal&-&SC (Direct)&0.5/0.98&Y (Cross-coupling reaction/Interfacial synthesis)&\citep{kan2018interfacial}\\
C$_{13}$N$_{3}$&10.61&Hexagonal&-&SM&-/-&-&\citep{tan2021dirac}\\
C$_{18}$N$_{6}$&16.038&Hexagonal&-&SC (Direct)&2.20/3.33&Y (Cross-coupling reaction/Interfacial synthesis)&\citep{kan2018interfacial}\\
C$_{19}$N$_{3}$&12.22&Hexagonal&-&SM&-/-&-&\citep{tan2021dirac}\\
C$_{22}$N$_{4}$&8.52 (and 8.11)&Rectangular&-&SC (Quasi-direct)&1.14/-&-&\citep{li2015new}\\
C$_{36}$N$_{6}$&18.664&Hexagonal&-&SC (Direct)&1.10/1.55&Y (Cross-coupling reaction/Interfacial synthesis)&\citep{kan2018interfacial}\\
\hline
\hline
\end{tabular}}
\end{table*}
\label{tab:mono}

\begin{table*}[h!]
\centering
\label{structure-hetero}
\caption{Table containing the basic details of carbon nitride bilayers and heterostructures. Here SC, M, SM, NC stand for the semiconducting, metallic, semi-metallic, nodal cylindrical electronic nature of the materials respectively.}
\resizebox{\textwidth}{!}{\begin{tabular}{c c c c c c c c c c c c c c}
\hline
\hline
Structure & a$_0$ &Lattice mismatch ($\%$)& Interlayer distance & Electronic nature (Type)& Bandgap (Theory/Expt.)& Synthesis (Method)&Reference\\
\hline
\hline
Bilayer C$_2$N&8.31&-&3.08&SC (Direct)&1.58/-&-&\citep{dabsamut2022electric}\\
Bilayer TH-CN$_2$&4.18 (and 5.78)&-&5.18-5.44&SC (Indirect)&4.48 and 4.53/-&-&\citep{wei2021new}\\
Bilayer C$_3$N&7.12&-&3.30&SC (Indirect)&(0.30-1.21)/(0.40±0.04)-(0.85 ± 0.03)&Y (Hydrothermal)&\citep{wei2021bandgap}\\
Bilayer H-C$_3$N$_2$&12.29&-&2.99-3.62&SC (Direct)&0.35-0.49&-&\citep{cai2023c}\\
Bilayer T-C$_3$N$_2$&8.63&-&2.71-3.40&SM and SC (Indirect)&(0.1-0.22)/-&-&\citep{cai2023c}\\
Bilayer g-C$_3$N$_4$&7.12&-&3.30&SC (Direct)&1.70/2.60&Y (Molecular composite precursors)&\citep{niu2019photocatalytic,dong2013situ}\\
Bilayer C$_9$N$_4$&9.64&-&3.45&NC&-/-&-&\citep{chen2018prediction}\\
C$_2$N/C$_3$N&8.41 (and 19.20)&-&Coplanar&M&-/-&-&\citep{xu2018design}\\
C$_2$N/C$_6$N$_6$&14.32&1.18&3.10&SC (Indirect)&2.00&-&\citep{mukherjee2021designing}\\
g-C$_3$N$_4$/g-C$_6$N$_6$&7.13&0.18&2.63&SC (Indirect)&2.60/-&-&\citep{liang2016photocatalytic}\\
g-C$_3$N$_4$/g-CN&7.10&0.8&3.41&SC (Indirect)&2.42/-&-&\citep{he2021constructing}\\
g-C$_3$N$_2$/g-CN&7.11&0.3&2.94 (and 3.42)&SC (Indirect)&0.86/-&-&\citep{yuan2024design}\\
\hline
\hline
\end{tabular}}
\end{table*}
\label{tab:bilayer}

\subsection{\textbf{Stacked bilayers and heterostructures:}}Layered stacking or lateral interfacing of atomic monolayers has opened up unprecedented opportunities to engineer 2D heteromaterials. Creating composites of two or more semiconductors (SCs) can often enhance charge separation, thereby improving catalytic efficiency. In typical composite materials, commonly referred as heterostructures, the two types of charge carriers are distributed across different components. This separation slows down electron-hole recombination and introduces a unique donor-acceptor band offset. Depending on the characteristics of this band offset and the direction of electron and/or hole transfer, these composite materials are categorized into three subtypes. For carbon nitrides, the last few years have been quite intriguing as far as the fabricated heterostructures are concerned and the rich outcomes have shown their promise. For example, the hydrogen evolution reaction (HER) via the electrocatalytic reduction of water has become a promising method for a green energy supply in the future. However, the carbon-based metal-free electrocatalysts show poor activity. In 2018, Xu and co-workers reported an in-plane heterostructure of two synthesized monolayers of C$_2$N and C$_3$N which showcases a directional transfer of electrons from C$_3$N to C$_2$N under the built-in potential and projects its potency as a metal-free photocatalyst \citep{xu2018design}. Furthermore, g-C$_3$N$_4$-based type-II heterojunction photocatalysts have shown remarkable improvements in photocatalytic activity due to the effective spatial separation of electron-hole pairs enabled by the band alignment between two semiconductors. Their catalytic performance has been significantly boosted by employing the Z-scheme mechanism \citep{liao2021emerging}. This Z-scheme plays a vital role in photocatalytic reactions that require both reducing and oxidizing capabilities. For instance, in photocatalytic water splitting, light energy drives the separation of water into hydrogen and oxygen. The Z-scheme facilitates efficient separation and transfer of photogenerated charge carriers between two different semiconductors, enabling the simultaneous reduction of protons to hydrogen and oxidation of water to oxygen. This mechanism is crucial for designing efficient photocatalytic systems for applications such as solar fuel production and environmental remediation as we have discussed in a later section of heterostructures. The combination of ultrathin carbon nitride nanostructures, strong interfacial interactions, and staggered band alignment has facilitated a Z-scheme pathway for efficient charge-carrier separation and transfer \citep{zhao2021boron}, achieving a solar-to-hydrogen efficiency of 1.16\% under one-sun illumination. In 2015, Akple \etal reported the synthesis of WS$_2$-graphitic carbon nitride (g-C$_3$N$_4$) composites using WO$_3$ and thiourea as precursors in a gas–solid reaction. Varying amounts of WS$_2$ were loaded onto g-C$_3$N$_4$ to create heterostructures, which exhibited improved photocatalytic activity for H$_2$ production under visible light illumination \citep{akple2015enhanced}. In 2021, theoretical research demonstrated that an electric field is generated from the (C-doped) TiO$_2$ (101) surface toward the (B-doped) g-C$_3$N$_4$ monolayer in pristine, C-doped, and B-doped g-C$_3$N$_4$/TiO$_2$ heterostructures. Additionally, a higher band-edge potential was observed on the (C-doped) TiO$_2$ (101) surface compared to the (B-doped) g-C$_3$N$_4$ monolayer. Consequently, the pristine (2.591 eV), C-doped (2.663 eV), and B-doped (2.339 eV) g-C$_3$N$_4$/TiO$_2$ heterostructures were identified as Z-scheme systems, which enhance charge separation while preserving significant redox capabilities \citep{lin2021enhanced}. More recently, the exploration of potential systems based on two-dimensional (2D) heterostructures composed of carbon, nitrogen, or similar main group elements with special emphasis on the dynamics of excited charge carrier transfer and recombination processes is provided by Ghosh \etal \citep{ghosh2024emergence}, which are crucial for developing efficient	photocatalytic systems for overall water splitting. A recent report shows a critical first-principle investigation of g-GeC/MoSe$_2$ van der Waals heterostructure which follows the Z-scheme photocatalytic mechanism and demonstrates enhanced light-harvesting efficiency across both the visible and ultraviolet wavelength ranges. In the same footing, a type-II vdW heterostructure of C$_2$N/C$_6$N$_6$ allows solar energy harvesting in the visible spectrum for water splitting \citep{mukherjee2021designing}. As far as the experimental realizations are concerned, in 2013, Dong \etal developed a facile in-situ method to form a g-C$_3$N$_4$/g-C$_3$N$_4$ metal-free heterojunction with molecular composite precursors to facilitate the charge separation \citep{dong2013situ}. For the process, 6 gm of thiourea and 6 gm of urea were diffused with 30 mL water in an alumina crucible. The solution was dried at a temperature of 60°C overnight to have the molecular composite precursors. These precursors in an alumina crucible were heated at 550 °C at a rate of 15°C per minute in a muffle furnace and kept for 2 hr. After the reaction, the alumina crucible was brought to room temperature and the resultant was collected for further utilization. An enhancement in the visible-light-absorption than a single layer is achieved by a bilayer g-C$_3$N$_4$ , and the calculated optical absorption threshold is significantly shifted downward by 0.8 eV, which is reported to be induced by the interlayer coupling \citep{wu2012visible}.

In 2021, Wei \etal reported the successful synthesis of C$_3$N bilayers where different proportions of 2,3-diaminophenazine (DAP) precursor and time were used to grow these structures \citep{wei2021bandgap}. The stacking orders can be tuned effectively by changing the reaction time of the hydrothermal treatment of DAP, and a higher growth time results in the formation of AA'(and AB') stackings. The DAP aqueous solution of 300.0 mL and 3.0 mM (200.0 mL, 2.0 mM) was added into a 100 mL PPL-lined stainless steel autoclave, heated at 200$^o$C for 72 hr (and 192 hr). The products were separated by centrifugation with the specifics of 10000 r per minute, 60 min, 10$^o$C. The electronic bandgap of 0.40±0.04 eV (and 0.85±0.03 eV) was obtained for C$_3$N bilayer with AA' (and AB') stacking - determined through a statistical analysis of 30 individual STS curves obtained on multiple bilayer samples. Interestingly, a new structure of nodal line, i.e., nodal cylinder, is found in momentum space for AA-stacking C$_9$N$_4$ \citep{chen2018prediction}. Table-3 here provides all the existing C$_x$N$_y$ hetero-bilayers with necessary details and references.

\section{Automation in material science}
Material science research has shown a paradigm shift whence the automated discovering of crystal structure was introduced. Different crystal search algorithms were proposed over time like particle swarm optimization in CALYPSO \cite{wang2010crystal} and other metaheuristic algorithms \cite{talatahari2021crystal}, evolutionary algorithm in USPEX \cite{glass2006uspex, oganov2011evolutionary}, simulated annealing in Sir2019 \cite{burla2015crystal}, EXPO2014 \cite{altomare2015expo}, metadynamics \cite{martovnak2005simulation, cavalli2015investigating}, and many others. Sometimes, a simple method such as an automated neighbour finding approach relying on the idea that increased covalent bonding nature within local substructure increases the stability may be proved enough for finding out new CN like structures \cite{villarreal2021metric}. With their merits and limitations these high-throughput studies can generate huge data on crystal structures and investigate their viability through stability check spanning all over the crystallographic space.

The application of ML in structure prediction is a relatively recent advancement. A ML based prediction and analysis framework, incorporating a symmetry based combinatorial crystal optimization program (SCCOP) and a feature additive attribution model, was developed to significantly reduce computational costs and extract property-related structural features \cite{li2023graph}. This approach was demonstrated practically on a 2D B-C-N system, showcasing its capability for high-throughput structural search and feature extraction. Initially, structures generated from 17 plane space groups were converted into crystal vectors using a graph neural network (GNN), followed by energy prediction. Bayesian optimization was employed to explore the structure's potential energy surface minimum. Subsequently, desired structures were optimized using ML-accelerated simulated annealing (SA), supplemented by a limited number of DFT calculations to achieve the lowest energy configuration.

Prediction of materials properties associated with the formation and stability such as formation enthalpy \cite{peterson2021materials,faber2015crystal}, phononic dispersion \cite{mortazavi2020exploring,cui2023machine} with at par accuracy of quantum mechanics calculations and much lower computational cost is becoming popular with time.
\begin{figure*}[ht]
	\centering
	\includegraphics[width=0.8\textwidth]{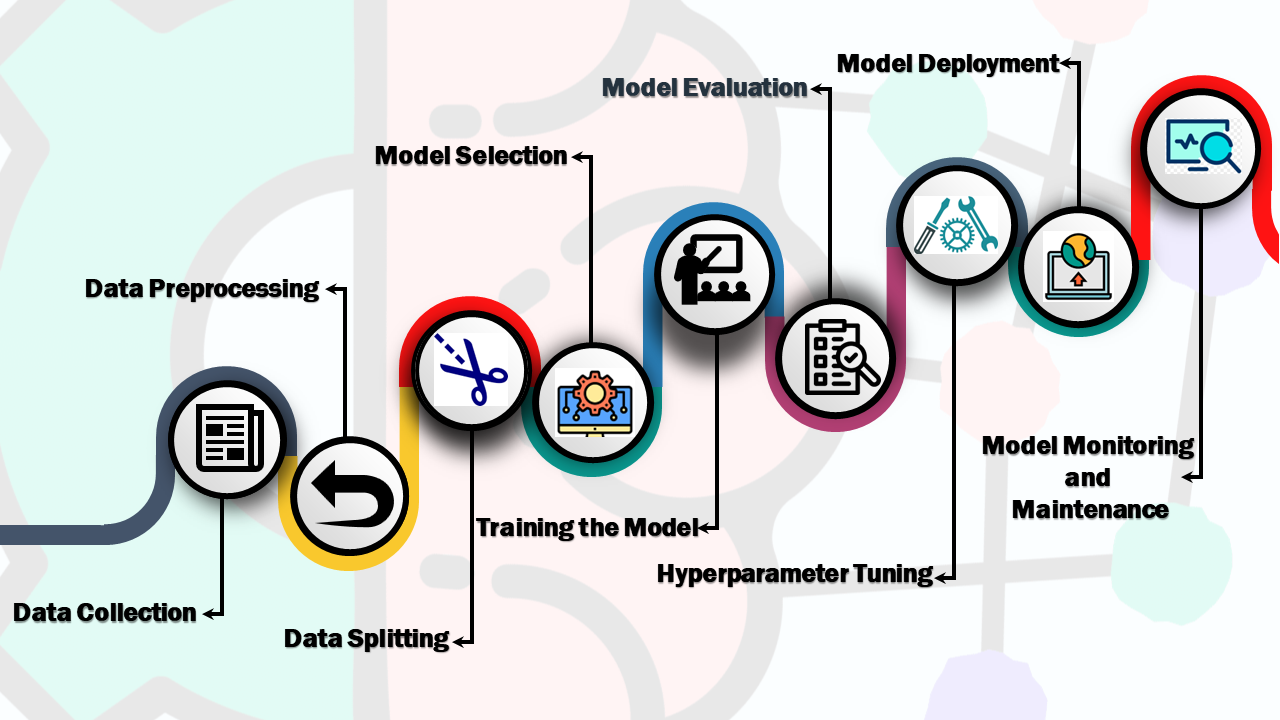}
	\caption{Different steps in Machine Learning process}
	\label{fig:ML_steps}
\end{figure*}
\begin{table*}[t!]
	\centering
	\small
	\caption{Platforms and Libraries for Machine Learning}	
	\begin{tabular}{p{3.5cm}p{13cm}}
		\toprule
		\textbf{Name} & \textbf{Description} \\
		\midrule
		Amazon SageMaker & Fully managed service by AWS for building, training, and deploying ML models at scale, supporting various frameworks. \\
		\addlinespace
		Azure ML & Cloud based service by Microsoft for building and deploying ML models, supporting automated ML and scalable computing resources. \\
		\addlinespace
		Caffe & Deep learning framework with expressive architecture and model scalability, suitable for image recognition tasks. \\
		\addlinespace
		caret & R package for data preprocessing, model training, and evaluation across various ML algorithms. \\
		\addlinespace
		CatBoost & Gradient boosting library by Yandex with advanced handling of categorical features for ML tasks. \\
		\addlinespace
		DataRobot & Automated ML platform for building and deploying ML models without extensive expertise. \\
		\addlinespace
		Dask-ML & Scalable ML library compatible with scikit-learn APIs, designed for parallel and distributed computing with Dask. \\
		\addlinespace
		H2O.ai & Open-source ML platform with scalable algorithms, AutoML, and support for multiple languages. \\
		\addlinespace
		IBM Watson Studio & Integrated environment for data scientists and developers to build and deploy AI models using various open-source tools and IBM services. \\
		\addlinespace
		KNIME & Open-source data analytics platform with visual programming interface for building data analysis workflows. \\
		\addlinespace
		LightGBM & Gradient boosting framework by Microsoft for high-performance ML tasks with low memory usage. \\
		\addlinespace
		MATLAB & Proprietary programming language and environment for numerical computing and ML. \\
		\addlinespace
		ML.NET & Open-source ML framework by Microsoft for building and deploying ML models in .NET applications. \\
		\addlinespace
		MLlib (Apache Spark) & Distributed ML library within Apache Spark for large-scale data processing and model training. \\
		\addlinespace
		MXNet & Open-source deep learning framework for flexible and efficient model building and deployment. \\
		\addlinespace
		Orange & Open-source tool for data visualization, analysis, and ML with a visual programming interface. \\
		\addlinespace
		PyTorch & Open-source ML framework by Facebook, known for flexibility and ease of use in deep learning research. \\
		\addlinespace
		RapidMiner & Integrated data science platform with visual workflow for data preparation, ML, and analytics. \\
		\addlinespace
		scikit-learn & Python library for ML tasks such as classification, regression, and clustering. \\
		\addlinespace
		Spark MLlib & Apache Spark's scalable ML library for distributed data processing and model training. \\
		\addlinespace
		TensorFlow & Open-source ML framework by Google, supporting deep learning and scalable model deployment. \\
		\addlinespace
		Theano & Python library for defining, optimizing, and evaluating mathematical expressions with support for deep learning. \\
		\addlinespace
		Weka & Java based collection of ML algorithms and tools for data mining tasks. \\
		\addlinespace
		XGBoost & Scalable and efficient gradient boosting library for ML tasks, known for high performance and accuracy. \\
		\bottomrule
	\end{tabular}
\end{table*}
\label{tab:ml_platforms}
\section{The Machine Learning Modelling}
The key of Machine Learning is data. Larger the dataset is, precise the predictions are. In human idea, gathering data is like gathering experiences. The standard of dataset depends on the volume as well as on the versatility. In material science the volume of data is increasing day by day due to the automated high-throughput discovery of materials properties. Different platforms as tabulated in Table-4 are readily available for throughput calculation. 
Datasets ideally have some input parameters and the corresponding outputs. The next step in ML algorithm is finding relevant input descriptors those can be mapped to the outputs. This step is called featurisation. Once the featurisation completed, a suitable ML model is built mapping the descriptors to the output. Then the model is tested to evaluate its predictive power. A brief description of the steps (see Figure-6) follows:

\subsection{Data collection and curation} 
Collecting data is the very first step in ML. Materials' data can be acquired either from experimental or from theoretical studies. Different databases are maintained, some of those are listed in Table-5. Now, the experiments are done in different conditions or the methods for characterising properties differ, so, differences in results is ubiquitous. If the data is checked before entering in databases, this variation remains. Similar situation is observed for computational data due to methodological variation. Besides the available databases, data can be acquired directly from the research articles through natural language processing, however, quality checking through human intervention is highly required till now.

Data collected from different sources may incur confusion. Data curation is an essential step to manage and maintain the data. It helps to improve the quality of the data through cleaning, inconsistency correction, and removing duplicate entries. In some cases the database may have some missing entries those are addressed in this step as well. 

\subsection{Featurisation}
The data has to be presented in a mathematical form understandable by computers. This involves expressing the characters of materials and its constituents (atoms, molecules, ligands, etc.) by numerical values. For example, in any solid formed by different elements, the mass number, number of electrons, valence electronic configuration, crystal structure, symmetries, coordination number of the atoms, etc. can be treated as the fingerprints or descriptors. These descriptors should be unique from each other and the number of descriptors should be optimum, not so much that the model overfits or not so less that the vital parameters are left aside. There are chances of having different descriptors with high correlation or descriptor with almost same value throughout. Feature selection is a technique to reduce the dimensionality of input parameters so that a smaller subset of the relevant features can be extracted through the removal of redundant, irrelevant, or noisy descriptors. This is done using different supervised or unsupervised techniques:

\subsubsection{Supervised techniques} In case of supervised learning the input data are labelled, which means every element is tagged with a correct classification. For example, if a number of figures of cubic and orthorhombic unit cells are provided and each of those are classified then the data is labelled. From this data, the machine may learn to identify a unit cell not in this dataset as cubic or orthorhombic. So, from the initial point the labelling is output oriented. Supervised techniques of feature selection are broadly put into three categories: filter based, wrapper based and embedded approaches.

In filter based strategy the correlation of data with each other is evaluated not considering the ML model to be used. This can be as simple of seeing the importance (importance score) of all the features by sorting, categorised as information-gain method. Another widely used technique is chi-squared-test which is a simple statistical tool to evaluate the correlation of different entities. Method using Fisher's criteria to rank the variables in descending order for determining most interesting feature is another widely used technique. If there are many missing values of any parameter that descriptor should be dropped. This is done by calculating the missing value ratio which is the number associated with each column found by the ratio of total number of missing values and the total number of observations. 

The wrapper based method  utilises specific machine learning to determine the best combination of features through evaluation and comparison with other combinations. On the basis of performance, descriptors are added starting from an empty set (forward selection) or subtracted from the set of all descriptors (backward selection), and in every step the model is trained again. Both of these are serial methods, so, all combinations are not tested. Exhaustive selection method evaluates all possible subsets of features to identify the best performer.

While there is no option of learning in filter methods, wrapper methods use separate learning algorithm for feature selection. In embedded method the feature selection embeds with the ML model algorithm itself and feature selection part can not be separated from the learning part. These are faster than wrapper based methods as the ML model is fitted only once. In regularisation technique a penalty term is added to the coefficients of different features so that for some of those the coefficients become zero, hence, the overfitting of the model can be avoided. On the other hand, decision tree based methods Random-forest or Gradient-boosting look at the outcomes from all the different nodes of decision trees. Through the tree building process, significance score for all the features are computed and thus sorted.

\subsubsection{Unsupervised techniques}Where supervised learning is goal oriented, unsupervised strategies use algorithms to find correlation in data without any explicit instruction. This works with unlabelled data. Principal component analysis is such an strategy that transforms the original correlated features into a set of orthogonal features while conserving the variance as much as possible. These principal components are essentially linear combinations of the original descriptors. Independent component analysis is a linear transformation method targeting the statistical independence of features as good as possible. Two random variables are statistically independent if the joint probability can be expressed as the multiplication of individual probabilities. Beside these, there are other strategies with higher complexity like Non-negative matrix factorization, Autoencoders, etc.

\subsection{Splitting data in training, validation and test subset} This is one of the crucial steps for ML before selection of model. The model is fit on a dataset called training dataset. The fitted model is then used to predict outputs on a second dataset named as the validation dataset to provide an unbiased evaluation of the trained model. The hyperparameters of the model is tuned depending on this evaluation and a final model is prepared. Finally, test dataset is used to evaluate the performance of the final model.

The simplest method to split the data is Random splitting which divides the dataset randomly into these three types. Random splitting works well on balanced data where all the types have enough examples in the dataset. For imbalanced datasets, to ensure the consistency in class distribution among subsets, stratified splitting technique is used. Now for sequential data (time-series data) random sampling can be chaotic, so, time-series splitting preserves the temporal order of data. All of these are one-shot methods. Cross validation method like K-fold cross validation produces “k” equally sized subset from the dataset. In each of k-step iteration, one of these work as a validation subset and the other (k-1) subsets as training subsets. For small dataset Leave-one-out cross validation is an effective method.  At an iteration step, one sample is taken as a test sample, and the remaining samples form the training set. The iteration is repeated for every sample. In cross validation methods the evaluation results are averaged over the number of iterations to evaluate the performance of ML model. 

\subsection{Model selection and training} 
\begin{figure*}[t!]
	\centering
	\includegraphics[width=0.8\textwidth]{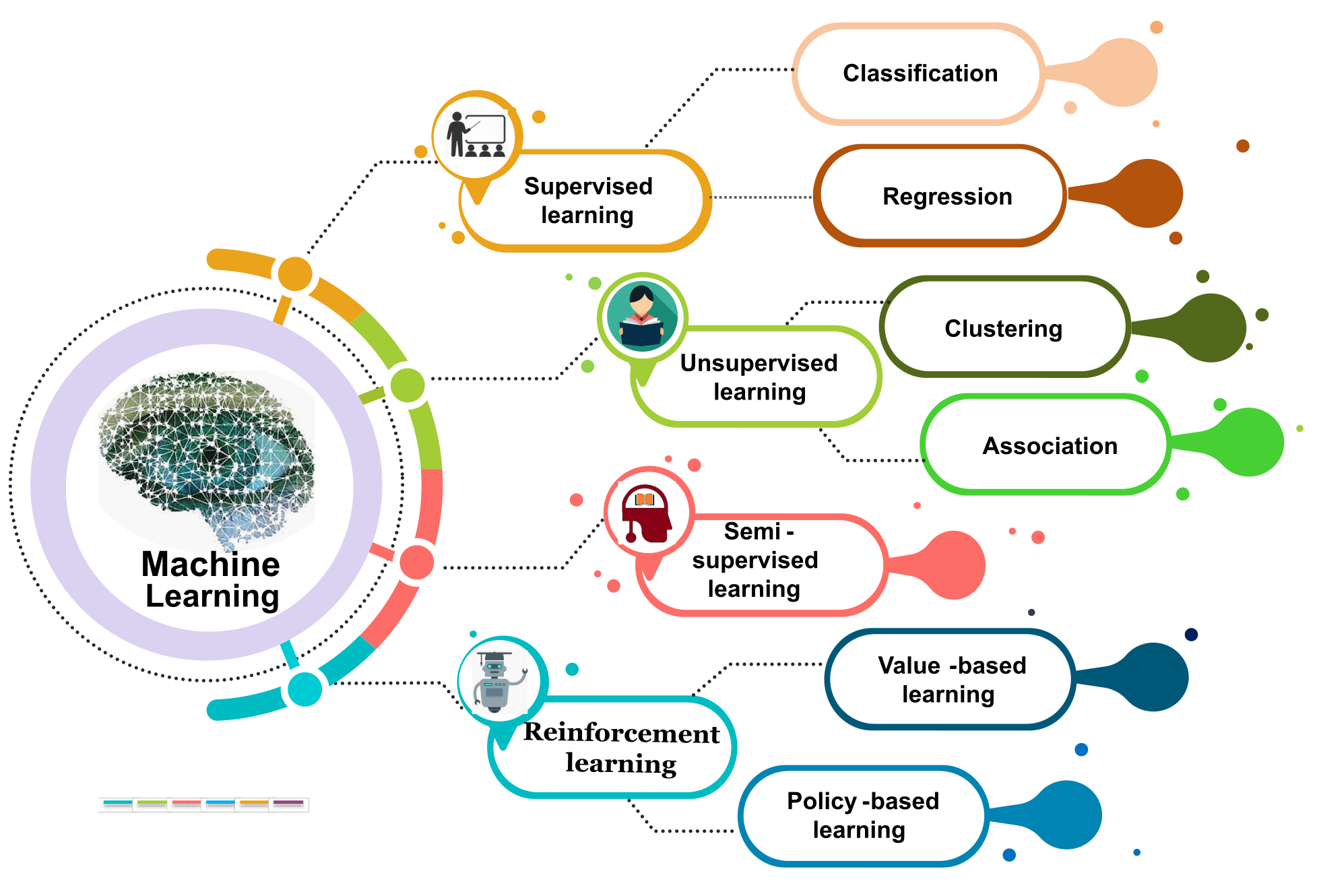}
	\caption{Classification of Machine Learning models}
	\label{fig:ML_classification}
\end{figure*}

This the most important part. Choice of ML model is very much problem dependent as well as performance oriented. The ML models are broadly classified in three classes (see, Figure--\ref{fig:ML_classification}):
\begin{itemize}
	\item Supervised learning
	\item Unsupervised learning 
	\item Reinforcement learning
\end{itemize}
To estimate how well a particular ML model performs, loss function is calculated. Different loss functions are used:
\begin{itemize}
	\item Mean Squared Error (MSE): $$MSE = \frac{1}{n} \sum_{i}^{n}({y}_i-\hat{y}_i)^2$$ where, $\hat{y}_i$ and $y_i$ are the original and predicted output values.
	\item Mean Absolute Error (MAE): $$MAE = \frac{1}{n} \sum_{i}^{n}|{y}_i-\hat{y}_i|$$
	\item Huber Loss: Hybrids MSE and MAE in single loss function:
	\begin{equation*}
		L_{\delta,i}(\hat{y_i},y_i) = 
		\begin{cases}
			\frac{1}{2} (\hat{y_i}-y_i)^2 & \text{if $|\hat{y_i}-y_i| \le \delta$}\\
			\delta (|\hat{y_i}-y_i|-\frac{\delta}{2}) & \text{otherwise}
		\end{cases}       
	\end{equation*}
	\item Hinge loss: This is another loss function for binary classification, where, two classes of data points labelled as $+1$ and $-1$ are meant to be separated.
	$$L_i = max [0, (1-\hat{y}_i).y_i]$$
	where, $\hat{y}_i$ is the actual class ($-1/1$) and $y_i$ is the classifier predicted value for the data point.
	\item Binary cross entropy loss/ Log loss: For classification problem the question arises, whether the output is this (let, put numerical value $y=1$ for YES) or not ($y=0$ for NOT). If $p(y_i)$ is the probability of finding YES at any data point $i$ then 
	$$ L = -\frac{1}{n} \sum_{i}^{n} [y_i .log(p(y_i)) + (1-y_i).log(1-p(y_i))]$$
	\item The coefficient of determination: The coefficient of determination, \( R^2 \), is a statistical measure that indicates how well the regression predictions approximate the real data points. It is defined as:
	\[
	R^2 = 1 - \frac{SS_{\text{res}}}{SS_{\text{tot}}}
	\]
	where \( SS_{\text{res}} \) is the sum of squares of residuals (the differences between the observed and predicted values) and \( SS_{\text{tot}} \) is the total sum of squares (the sum of squared differences between the observed values and their mean).	
	\( R^2 \) ranges from 0 to 1, where, \( R^2 = 1 \) indicates that the model perfectly predicts the dependent variable using the independent variables, and \( R^2 = 0 \) indicates that the model does not explain any of the variance in the dependent variable around its mean.
	
\end{itemize}

\begin{figure*}[ht]
	\centering
	\includegraphics[width=\textwidth]{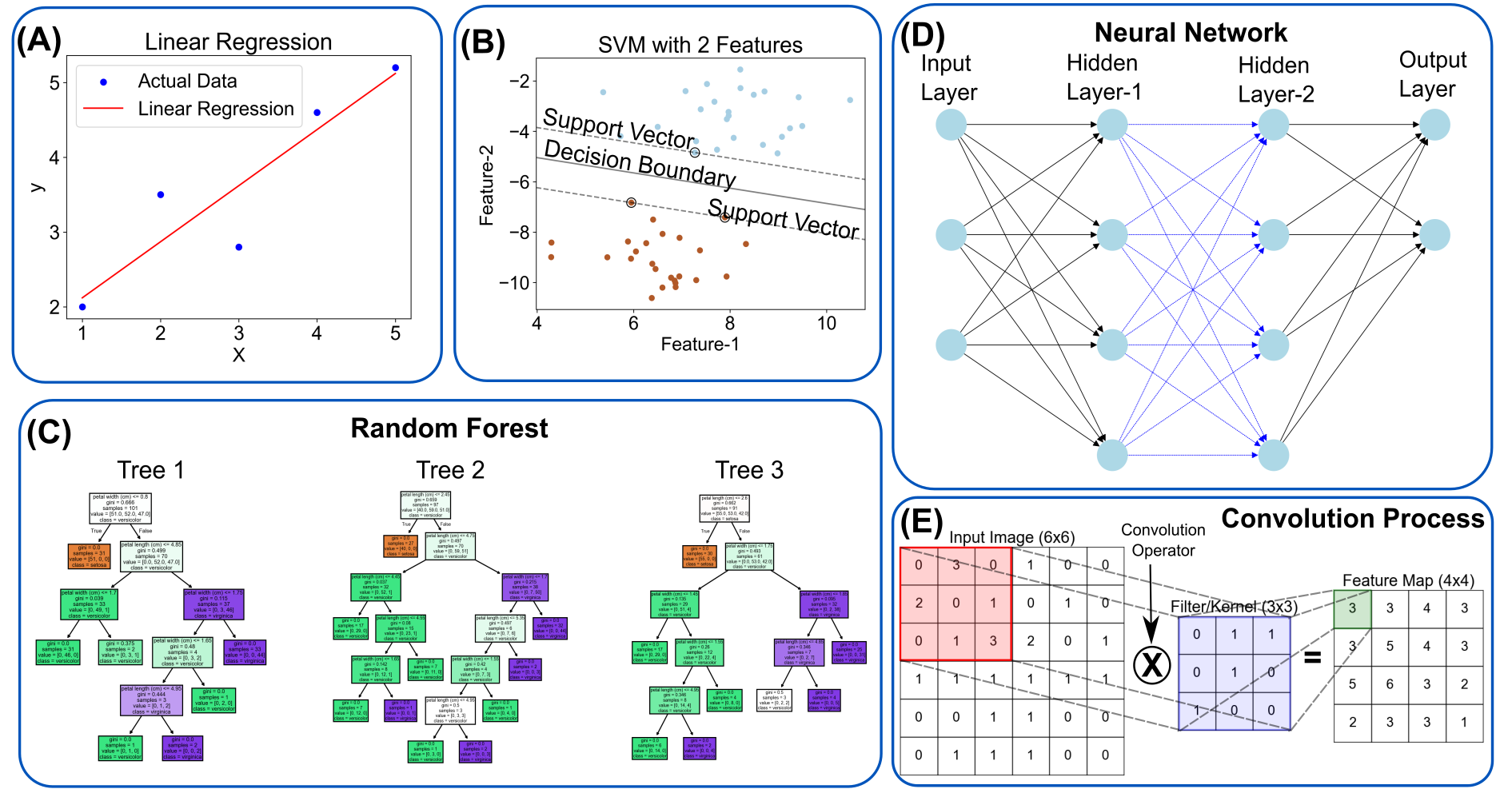}
	\caption{Different Machine Learning models. Linear Regression, Support Vector Machine with two features. A Random Forest with three decision-trees. Artificial Neural Network with more than one hidden layers identified as Deep Neural Network and the convolution process, precursor to Neural Network in convolutional Neural Network.}
	\label{fig:ML_model}
\end{figure*}

\subsubsection{Supervised ML models} As discussed earlier supervised technique works with labled data and essentially goal oriented. Further subdivisions of supervised learning are: Classification and Regression.\\In regression, the output parameter is a continuous variable, while classification outputs discrete variables. For example, categorizing solids as metals or insulators is a classification problem, whereas, predicting the bandgap of any solid from different input variables is a regression problem.

\subsubsection*{Regression:} Regression analysis is a statistical method used to model the relationship between one or more independent variables and a dependent variable. It seeks to identify and quantify the impact of these variables on the outcome, facilitating prediction and understanding of the underlying data patterns. There are numerous types of regression methods used in ML (see, Table- \ref{tab:regression_pros_cons} for their strong and weak points), for example:

\begin{itemize}[]
	\item {\itshape Linear Regression (LR):} The linear relationship between the dependent output variable (y) and one or more independent input features (x's) by fitting a straight line $y=\sum_{i}m_ix_i+c$ is established through this simplest type of regression method.
	
	\item {\itshape Multiple Linear Regression (MLR):} Multiple Linear Regression (MLR) is a statistical method used to model the relationship between multiple independent variables and a continuous dependent variable. It assumes a linear relationship between predictors and the target variable, making it straightforward to interpret. MLR estimates the coefficients of each variable to minimize the sum of squared differences between predicted and observed values. While simple and efficient, MLR may struggle with non-linear relationships and interactions between predictors.
	
	\item {\itshape k-Nearest Neighbors Regression (KNR):} It is a non-parametric machine learning algorithm used for regression tasks. It predicts the value of a new data point by averaging the values of its k nearest neighbors in the training dataset. The 'k' in KNR represents the number of neighbors considered for prediction, which is typically chosen based on cross-validation or other model selection techniques. KNR does not make any assumptions about the underlying data distribution and can capture complex relationships between input and output variables. However, its performance can be sensitive to the choice of k and the distance metric used to measure similarity between data points.
	
	\item {\itshape Ridge Regression (RR):} Ridge Regression is a linear regression technique that incorporates regularization to address multicollinearity and improve model generalization. It adds a penalty term to the standard least squares objective function, which is proportional to the square of the magnitude of the coefficients. This penalty term (controlled by a regularization parameter, typically denoted as \(\alpha\)) shrinks the coefficients towards zero, effectively reducing their variance and mitigating the influence of correlated predictors. RR is particularly useful when the number of predictors (features) is large, or when predictors are highly correlated. By imposing regularization, RR helps to stabilize the model and improve its performance on new, unseen data. The optimal value of \(\alpha\) is usually determined through cross validation.

	\item {\itshape Kernel Ridge Regression (KRR):} It is a supervised learning algorithm used for regression tasks. It extends RR by using the kernel trick, which allows it to handle non linear relationships between input variables and the target variable. KRR minimizes a regularized objective function that includes both the squared error loss and a penalty term that encourages smoothness in the predictions. The choice of kernel function (such as linear, polynomial, Gaussian radial basis function, etc.) determines the type of non-linear transformation applied to the data. KRR is particularly useful when the relationship between input and output variables is complex and cannot be effectively modeled by linear regression. However, it requires tuning of hyperparameters such as the regularization parameter and the kernel parameters to achieve optimal performance.
	
	\item {\itshape Gradient Boosting Regression (GBR):} Gradient Boosting Regression (GBR) is an ensemble learning technique that sequentially combines weak learners, typically decision trees, to build a predictive model. It works by fitting new models to the residuals of the previous predictions, thereby minimizing the error. GBR is known for its ability to capture complex interactions and non-linear relationships in data, making it a powerful tool for predictive modeling. However, it can be computationally intensive and prone to overfitting if not carefully tuned.
	
	\item {\itshape Extreme Gradient Boosting Regression (XGBoost):} Extreme Gradient Boosting Regression (XGBoost) is an advanced implementation of gradient boosting (GBR) that has gained popularity for its speed and performance. It uses a regularized objective function and parallel computing to optimize model training and prediction. XGBoost excels in handling large datasets, missing values, and complex interactions between variables. Despite its effectiveness, XGBoost may require careful tuning of parameters to prevent overfitting and achieve optimal performance.
	
	\item {\itshape Support Vector Regression (SVR):} Support Vector Regression (SVR) is a regression technique based on Support Vector Machines (SVMs), which find the optimal hyperplane that best fits the data while minimizing prediction errors. SVR is particularly effective in high dimensional spaces and can handle non linear relationships through the use of kernel functions. However, SVR's performance heavily depends on the choice of kernel and regularization parameters, which can be challenging to optimize. It is also sensitive to outliers in the data, requiring careful preprocessing steps.
	
	\item {\itshape Least Absolute Shrinkage and Selection Operator (LASSO):} LASSO regression is a linear regression technique that incorporates regularization to improve the model's prediction accuracy and interpretability. It penalizes the absolute size of the regression coefficients, forcing some coefficients to shrink towards zero. This encourages sparsity and feature selection, making Lasso particularly useful when dealing with datasets with many correlated variables or when aiming to identify the most important predictors. Adjusting the regularization parameter allows for control over the degree of regularization applied, balancing model complexity and predictive performance.
\end{itemize}

\subsubsection*{Decision Trees (Classification/Regression)}
A decision tree is a flowchart like tree structure. It is used to make predictions or decisions. In this tree structure the nodes represent tests, branches represent the different outcomes, and leaf nodes represent the predictions.

\subsubsection*{Random Forests (Classification/Regression)}
In Random Forest (RF) strategy multiple decision trees are built to reach a single result. This is an ensemble method made up of a set of these trees and their predictions are aggregated to recognize the result with majority of votes.

\subsubsection*{Extra-Trees (Classification/Regression)}
Extra Trees, or Extremely Randomized Trees, is an ensemble learning method based on decision trees, similar to RFs but with additional randomness in the selection of splitting points. This approach further diversifies the trees by selecting random thresholds for node splitting, aiming to improve generalization and robustness by reducing variance and potential overfitting in regression tasks.

\subsubsection*{Support Vector Machine(Classification)}
Support Vector Machine is used mostly for linear or non-linear classification, however, it is also used for regression.  The objective of SVM is to determine the optimal hyperplane that can isolate the data points in separate classes. If there are two distinct classes, the hyperplane is just a line, for three classes two planes can separate. In general, in SVM model a data point is viewed as a n-dimensional vector and it tries to find a (n-1) dimensional hyperplane to separate those points. Though the original proposal was a linear hyperplane, later non-linear SVM was proposed \cite{aizerman1964theoretical}.

\subsubsection*{Naïve Baye's(Classification)}
The Baye's theorem on conditional probability is expressed as
$$P(A|B) = \frac{P(B|A)P(A)}{P(B)}$$
where, $P(A|B)$ is the conditional probability of event$-A$ when event$-B$ is true, $P(B|A)$ is the conditional probability of event$-B$ when event$-A$ is true, and $P(A)$, $P(B)$ are the probabilities of $A$ and $B$ without any given condition. For a class variable $y$ and dependent features $x_i$'s, this looks like
$$P(y|x_1,x_2,..., x_n) = \frac{P(x_1,x_2,..., x_n|y)P(y)}{P(x_1,x_2,..., x_n)}$$

Now, the "naive" conditional independence approximation assumes that all the features are mutually independent, so that:
$$P(x_i|y,x_1,x_2,...,x_{i-1},x_{i+1},...,x_n) = P(x_i|y)$$
$$\implies P(y|x_1,x_2,..., x_n) = \frac{P(y)\prod_{i=1}^{n} P(x_i | y)}{P(x_1,x_2,..., x_n)}$$
For a given set classes, the classification output is found simply as:
$$y = \arg\max_{k} P(y_k) \prod_{i=1}^{n} P(x_i | y_k)$$
There are different types of naive Baye's algorithm: Gaussian naive Baye's, Multinomial naive Baye's, Bernoulli naive Baye's, etc.

\subsubsection*{Gaussian Process Regression (GPR)}
Gaussian Process Regression (GPR) is a probabilistic supervised learning method used for regression tasks. It models the relationship between input variables and the target variable as a Gaussian process, which is a collection of random variables indexed by the input space. GPR assumes a prior over functions, typically Gaussian, and updates this prior based on observed data to obtain a posterior distribution over functions that best fit the data. Predictions are made by computing the mean and variance (or uncertainty) of the posterior distribution at new input points. GPR is flexible in capturing complex patterns in data without assuming a specific functional form, making it suitable for tasks where the underlying relationship is not well understood or may exhibit non-linearities. However, its computational complexity increases with the size of the dataset, requiring efficient approximation methods for scalability.

\subsubsection*{Artificial Neural Networks:}
In brain, neurons have very complex network. The dendrites of neurons take the input and axon transfer the output to other neurons (see Figure-\ref{fig:ML_model}). Synapses handle the transmission of nervous impulses between neurons or neurons to cells. Motivated by this, the artificial neural networks (ANN) have input nodes, output nodes and weights to tune the transmission of data from output nodes to input nodes of other neurons. In ANN there is an input layer, an output layer and one or more hidden layers in between.

The most basic ANN is unidirectional feedforward neural network. The data enters through the input layer and go out through the output layer,. The hidden layers between input and output may or may not exist. Through the backpropagation technique the weights between nodes can be adjusted depending on the error calculation. A comprehensive list of different ANN architectures, their strong and weak points are tabulated in Table-\ref{tab:ann_pros_cons}.


If a series of layers is used in feedforward ANN, then generally it is called Deep Learning. In deep neural network (DNN) the useful feature are automatically extracted from the data. It can model non-linear relationships as well.

In mathematics convolution is defined as a way of combining two functions into third function. In convolutional ANN (CNN) a set of learnable filters is used on data to to extract features. This is done convolution layer as depicted in Figure--\ref{fig:ML_model}. This is essentially a DNN.

A feedback neural network allows signals to use feedback paths for travelling in both directions.  This type of network contains loops, thus a nonlinear dynamical system is produced. Recurrent neural network (RNN) is a prominent variety of feedback neural network. RNNs are designed for sequential data processing, capable of capturing dependencies and patterns over time through recurrent connections. They are widely used in tasks such as natural language processing, time series prediction, and speech recognition, where temporal dynamics play a crucial role. 

General Regression Neural Networks (GRNN) are a type of neural network architecture utilized primarily for regression tasks. They employ radial basis functions in their hidden layer neurons to approximate continuous functions based on input data. GRNNs are known for their capability to model complex and nonlinear relationships between input and output variables effectively. Unlike traditional neural networks, GRNNs adopt a memory based learning approach, storing all training samples in memory for rapid prediction during inference. This architecture makes GRNNs suitable for applications requiring accurate regression predictions, where the underlying data relationships are intricate and may not conform to linear patterns.

\textit{Activation functions: }Activation functions are crucial components in artificial neural networks, responsible for introducing non-linearity into the network's output. They operate on the weighted sum of inputs to a neuron and determine its output. Each activation function serves a specific purpose, facilitating the network's ability to model complex relationships within data.

The linear activation function, denoted as \texttt{purelin}, is defined as:
\[ \text{purelin}(x) = x \]
This activation function simply outputs the input \( x \) without applying any transformation. It is typically used in the output layer of neural networks for regression tasks, where the network needs to predict continuous values. Purelin preserves the linearity of the input data and produces outputs that are directly proportional to the input, making it straightforward for interpretation and computation in regression scenarios.

The logistic sigmoid activation function, often referred to as \texttt{logsig}, is defined as:
\[ \text{logsig}(x) = \frac{1}{1 + e^{-x}} \]
It maps the input \( x \) to a value between 0 and 1, making it suitable for binary classification tasks or as a squashing function to normalize outputs in neural networks. This activation introduces non-linearity, enabling the network to learn complex patterns in the data by transforming the input into a probability-like output.

Additionally, the rectified linear unit (ReLU) is popular for hidden layers due to its computational efficiency and ability to handle vanishing gradient problems.

\subsubsection{Unupervised ML models}
Unsupervised learning works with unlabeled data. Generally, three types of jobs are done using unsupervised learning:
\begin{itemize}
	\item Clustering: It is a method of grouping the data based on the similarities. Segmentation of  data into different groups enables analysis on each of those datasets to identify the inherent patterns.
	\item Dimension reduction: The number of features is called the dimensionality. For address overfitting issue and reduce the unneeded complexity of the model, dimension reduction is done.
	\item Association rule learning: The dependency of one item on another in a dataset is determined to identify the relations among variables so that any indication can be extracted. For example, if there are data on bandgaps and work functions or HOMO-LUMO positions in a dataset, it is likely that the article is targeted to find the catalytic merit of the material.
\end{itemize}

\subsubsection{Reinforcement Learning}
While in supervised learning labelled data is needed, reinforcement learning doesn't need that. Even it doesn't use unlabelled dataset as needed in unsupervised learning. It learns by its experience through the feedbacks and not intended in discovering a relationship in a dataset. Through trial and error, it tunes itself in accordance to the  environment, thus creates new datasets in each trial. As, deep learning method uses unstructured data, it is natural for almost all reinforcement learning strategies to use ANN in some extent. Depending on approach, the reinforcement learning can be categorised in two sub-categories:
\begin{itemize}
	\item Value based learning: The value based methods are meant to learn an optimal value function. The future reward of taking a step is anticipated for the present state.
	\item Policy based learning: The policy based methods are meant to learn an optimal policy directly. The model is built and updated without calculating any value function.
\end{itemize}

\begin{table}[]
	\centering
	\small 
	\resizebox{\textwidth}{!}{\begin{tabular}{p{3.5cm}p{7cm}p{6cm}}
			\toprule
			\hline
			\textbf{Model} & \textbf{Pros} & \textbf{Cons} \\
			\hline
			Linear Regression & 
			Simple and interpretable \newline
			Computationally efficient \newline
			Well-established statistical metrics \newline
			Good for linear relationships & 
			Assumes linearity \newline
			Sensitive to outliers \newline
			Can’t handle complex/non-linear patterns \newline
			Assumes normality of residuals \\
			\hline
			Ridge Regression & 
			Reduces overfitting with regularization \newline
			Works well with multicollinearity \newline
			Handles high-dimensional data effectively & 
			Limited interpretability \newline
			Cannot perform feature selection \newline
			Regularization parameter needs tuning \\
			\hline
			Lasso Regression & 
			Performs feature selection (shrinks coefficients to zero) \newline
			Reduces overfitting \newline
			Works well with sparse data & 
			Can miss important variables \newline
			Introduces bias \newline
			Regularization parameter needs tuning \\
			\hline
			Elastic Net & 
			Combines Ridge and Lasso strengths \newline
			Good for high-dimensional and correlated data \newline
			Performs both regularization and feature selection & 
			More complex to tune (two hyperparameters) \newline
			Less interpretable than Lasso alone \\
			\hline
			Polynomial Regression & 
			Models non-linear relationships \newline
			Flexible and adaptable & 
			Risk of overfitting \newline
			Higher complexity with degree of polynomial \newline
			Can be hard to interpret \\
			\hline
			Support Vector Regression (SVR) & 
			Works well in high-dimensional spaces \newline
			Robust to overfitting \newline
			Handles non-linear relationships via kernels & 
			Computationally expensive \newline
			Requires careful hyperparameter tuning \newline
			Less interpretable \\
			\hline
			Decision Tree Regression & 
			Easy to interpret \newline
			Handles non-linear data \newline
			No need for scaling or transformation \newline
			Simple visualization of decisions & 
			Prone to overfitting \newline
			Sensitive to small data changes \newline
			Instability with complex data \\
			\hline
			Random Forest Regression & 
			Robust to overfitting \newline
			Handles non-linearity \newline
			Can handle missing data \newline
			Good feature importance insights & 
			Less interpretable \newline
			Computationally intensive \newline
			May overfit if not properly tuned \\
			\hline
			Gradient Boosting (XGBoost, LightGBM) & 
			High predictive power \newline
			Handles complex non-linear relationships \newline
			Flexible with loss functions \newline
			Good feature importance & 
			Computationally expensive \newline
			Sensitive to hyperparameters \newline
			Risk of overfitting if not tuned \\
			\hline
			K-Nearest Neighbors (KNN) Regression & 
			Simple and intuitive \newline
			No assumptions about data distribution \newline
			Flexible and non-parametric & 
			Slow at prediction time \newline
			Sensitive to irrelevant features \newline
			Needs feature scaling \\
			\bottomrule
	\end{tabular}}		
	\caption{Pros and Cons of different regression models.}
	\label{tab:regression_pros_cons}
\end{table}

\begin{table}[]
	\centering
	\small 
	\resizebox{\textwidth}{!}{\begin{tabular}{p{3.5cm}p{5cm}p{5cm}}
			\hline
			\textbf{Model} & \textbf{Pros} & \textbf{Cons} \\
			\hline
			\textbf{Feedforward Neural Networks (FNN) / MLP} & 
			Simple architecture \newline
			Universal approximation \newline
			Versatile for regression & 
			Prone to overfitting \newline
			Slow training \newline
			No temporal memory \\
			\hline
			\textbf{Convolutional Neural Networks (CNNs)} & 
			Effective for images \newline
			Parameter sharing \newline
			Hierarchical feature learning & 
			Requires large datasets \newline
			Computationally expensive \newline
			Fixed input size \\
			\hline
			\textbf{Recurrent Neural Networks (RNNs)} & 
			Good for sequential data \newline
			Handles temporal dependencies \newline
			Flexible input length & 
			Vanishing/exploding gradients \newline
			Slow training \newline
			Limited long-term memory \\
			\hline
			\textbf{Long Short-Term Memory (LSTM)} & 
			Solves vanishing gradient problem \newline
			Captures long-term dependencies \newline
			Good for time series & 
			Computationally expensive \newline
			Slow training \newline
			Not ideal for non-sequential tasks \\
			\hline
			\textbf{Gated Recurrent Units (GRUs)} & 
			Simplified LSTM \newline
			Effective for sequential data \newline
			Faster training & 
			Less memory retention \newline
			Less interpretable \newline
			Potentially lower performance than LSTM \\
			\hline
			\textbf{Autoencoders} & 
			Dimensionality reduction \newline
			Feature extraction \newline
			Generative capabilities (e.g., VAEs) & 
			Overfitting \newline
			Poor reconstruction quality \newline
			Limited interpretability \\
			\hline
			\textbf{Generative Adversarial Networks (GANs)} & 
			Powerful for generative tasks \newline
			Unsupervised learning \newline
			Creative applications (e.g., art generation) & 
			Training instability \newline
			Requires fine-tuning \newline
			Hard to evaluate generated data \\
			\hline
			\textbf{Transformer Networks} & 
			State-of-the-art performance \newline
			Parallelizable \newline
			Effective for sequential data & 
			Data-hungry \newline
			Computationally expensive \newline
			Memory intensive \\
			\hline
			\textbf{Capsule Networks (CapsNets)} & 
			Better generalization \newline
			Resistant to spatial transformations \newline
			Dynamic routing of information & 
			Unstable training \newline
			Limited practical benchmarks \newline
			Slow convergence \\
			\hline
			\textbf{Self-Organizing Maps (SOMs)} & 
			Unsupervised learning \newline
			Topological preservation \newline
			Interpretability & 
			Slow training \newline
			Limited for non-linear tasks \newline
			Hard to scale \\
			\hline
	\end{tabular}}	
	\caption{Pros and Cons of different ANN architectures.}
	\label{tab:ann_pros_cons}
\end{table}

\section{Generative Artificial Intelligence}
\begin{figure}[t!]
	\centering
	\includegraphics[width=16 cm, height=9cm]{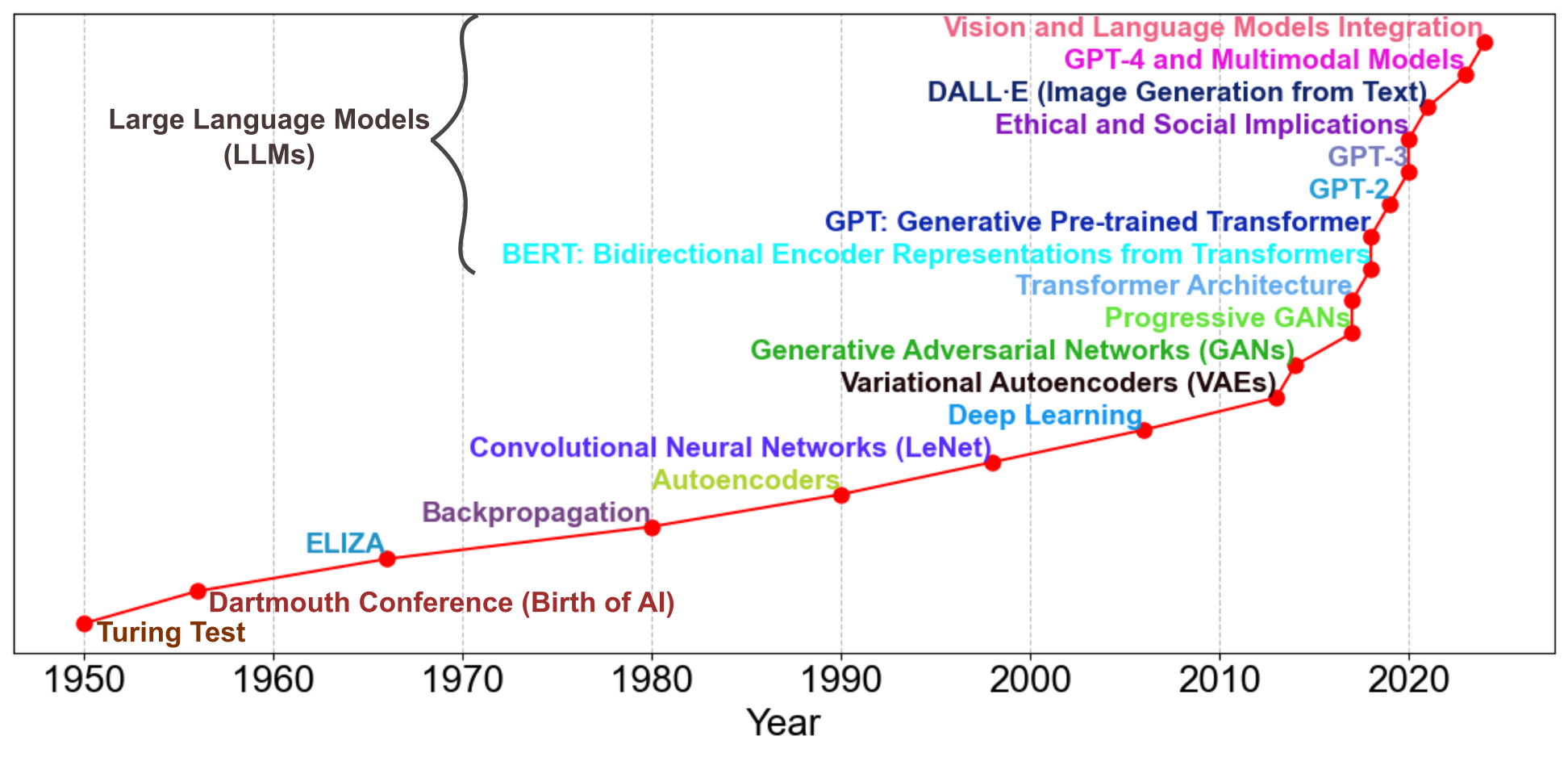}
	\caption{Timeline of Generative AI}
	\label{fig:timeline_vertical}
\end{figure}

While traditional ML models perform predefined tasks generative artificial intelligence (GAI) is a branch intended to generate new contents autonomously. Autoencoders emerged as the earlist form of generative models. Autoencoder is a type of ANN capable of learning network that is trained to compress input data effectively and regenerate it as output in unsupervised way. Variational autoencoders (VAE) introduced probabilistic modelling to autoencoders which allowed more flexibility in generating diverse outputs. Generative adversarial networks (GAN) revolutionized GAI by introducing a game-theoretical approach which resulted in highly realistic outputs. Both of these are very recent developments came after 2010. The timeline of the development of GAI is presented in Figure--\ref{fig:timeline_vertical}.

\subsection{Variational Autoencoders}
VAEs combine probabilistic modelling approach with ANN to represent complex dataset a lower-dimensional latent space. VAEs are applied in different generative tasks like image generation or natural language processing where generating novel and realistic data points is crucial.

\subsubsection*{Components}The different components of VAE are:
\begin{itemize}
	\item Encoder: The encoder maps input data points to a probabilistic distribution in the latent vector space. The input can be an image, or text sequence, or, similar unlabelled data.
	\item Latent space: The latent space is a vector space having dimension lower than the original input data depending on different encoding techniques. VAEs apply a probabilistic structure to it so that meaningful sampling and interpolation between points can be done.
	\item Decoder: The decoder takes information from the latent space and reconstructs an output that resembles the original input. By probabilistic sampling from the latent space VAEs can generate data points similar but not identical to the original data. This regeneration process makes VAE a powerful tools for generative tasks.
\end{itemize}

\subsection{Generative adversarial networks} 
GAN is made up of two ANNs, a discriminator and a generator. 
As training progresses, the generator improves its ability to produce more realistic data, guided by the feedback from the discriminator.
Conversely, the discriminator also improves over time, becoming more adept at distinguishing real from fake data.
\begin{itemize}
	\item Generator: The generator generates artificial data that resembles the input data. As input GAN takes latent vector or simply random noise. Its goal is to fool the discriminator by generating synthetic data indistinguishable from real data.
	\item Discriminator:  The discriminator evaluates whether a given data is real (from input) or fake (synthetic). Distinguishing between real and produced data is essentially a binary classification task.
\end{itemize}
The two components of GAN, the generator and discriminator, are trained in parallel and are intended to compete. Initially, the discriminator easily identifies the random noise generated by the generator as fake. With iterations, both the generator and the discriminator improve by taking competitor feedback. The training process continues until an equilibrium is reached where the generator can generate synthetic outputs that are realistic enough to fool the discriminator.
\begin{table*}[t!]
	\centering
	\caption{Online Platforms and Frameworks for Generative AI}
	\begin{tabular}{p{6cm}p{10cm}}
		\toprule
		{Platform/Framework} & \textbf{Description and Applications} \\ \midrule
		{OpenAI GPT} & Language model for text generation and understanding. Used in chatbots, content generation, translation. \\
		{DeepMind AlphaFold} & AI system predicting protein structures based on amino acid sequences. Applications in bioinformatics and drug discovery. \\
		{NVIDIA GANs (e.g., StyleGAN)} & Frameworks for generating high-quality images and manipulating visual content. Used in art generation, image synthesis, video creation. \\
		{TensorFlow} & Google's open-source framework for building and deploying ML models, including generative models like VAEs and GANs. \\
		{PyTorch} & Facebook's open-source framework known for its dynamic approach to neural networks. Used in research and production of generative models. \\
		{Hugging Face Transformers} & Library for working with transformers, including pre-trained models like BERT and GPT. Applications in text generation, sentiment analysis, and translation. \\
		{Google Cloud AI Platform} & Suite of AI services offering pre-trained models and tools for custom model development and deployment. \\
		{AWS AI Services} & Amazon's AI services providing pre-trained models and infrastructure for building custom AI solutions using TensorFlow, PyTorch, etc. \\
		{Microsoft Azure AI} & Cognitive services and ML tools for developing and deploying generative models, supporting custom AI solutions. \\
		\bottomrule
	\end{tabular}
	\label{tab:generative_ai}
\end{table*}
\begin{table*}[t!]
	\centering
	\small
	\resizebox{\textwidth}{!}{\begin{tabular}{p{4cm}p{12cm}p{5cm}}
			\toprule
			\textbf{Name} & \textbf{Short Description} & \textbf{Website} \\
			\midrule
			AFLOW \& AFlowLib & Automatic Flow for Materials Discovery, providing data and software. & \url{aflow.org} \url{aflowlib.org} \\
			AiiDA & Flexible and scalable informatics infrastructure to manage, preserve, and disseminate the computational data. & \url{aiida.net} \\
			Atomate & A high-level Python framework for automating computational workflows. & \url{atomate.org} \\
			C2DB & The Computational Crystallography Database of 2D materials. & \url{cmr.fysik.dtu.dk/c2db/c2db.html} \\
			ChemAI & An AI platform focusing chemical and materials discovery. & \url{acnetwork.nl/chemai} \\
			Citrine Informatics & Provides AI driven materials data and tools for materials discovery and optimization. & \url{citrine.io} \\
			DeepChem & An open-source deep--learning library applicable to chemistry and materials science. & \url{deepchem.io} \\
			Exabyte.io & A cloud based platform for computational materials science and chemistry simulations. & \url{exabyte.io} \\
			Granta MI & Materials information management software for managing and analysing materials data. & \url{grantami.com} \\
			HTEM DB & High Throughput Experimental Materials Database. & \url{htem.nrel.gov} \\
			JARVIS & Joint Automated Repository for Various Integrated Simulations for materials data and tools. & \url{jarvis.nist.gov} \\
			Khazana & A data repository for ML in chemistry and materials science. & \url{khazana.gatech.edu} \\
			Magpie & A ML package for materials data and prediction. & \url{bitbucket.org/wolverton/magpie} \\
			MAST-ML & An open-source tool for ML in materials science. & \url{mastml.org} \\
			Materials Cloud & Repository of open-access research data in computational materials science. & \url{materialscloud.org} \\
			Materials Data Facility & A repository and data resource that enables the publication, discovery, and access to materials data. & \url{materialsdatafacility.org} \\
			Materials Project & Provides open web based access to computed information on known and predicted materials. & \url{materialsproject.org} \\
			MatWeb & An online database for materials properties and ML models. & \url{matweb.com} \\
			Material Bank & A repository of materials data for use in ML and materials research. & \url{materialbank.org} \\
			Matbench & A benchmark suite for evaluating ML models in materials science. & \url{matbench.materialsproject.org} \\
			MatDL & Materials Data Lab, a platform for materials data management and analysis. & \url{matdl.org} \\
			Matminer & A library that contains tools to apply data mining and ML techniques to materials science. & \url{hackingmaterials.lbl.gov/matminer} \\
			MatSci ML & A platform focused on ML applications in materials science. & \url{github.com/IntelLabs/matsciml} \\
			MPContribs & A web platform to upload, share, and discuss contributions to the Materials Project database. & \url{mpcontribs.org} \\
			MPDS & Materials Platform for Data Science, a platform for materials data and tools. & \url{mpds.io} \\
			MDAnalysis & A Python library to analyse molecular dynamics simulations, providing data and tools for materials science. & \url{mdanalysis.org} \\
			NREL Materials Database & A repository of data on materials for energy applications. & \url{materials.nrel.gov} \\
			NOMAD & The Novel Materials Discovery (NOMAD) repository contains a vast amount of materials data and offers tools for their analysis. & \url{nomad-coe.eu} \\
			OMDB & Organic Materials Database provides access to electronic structure properties of previously synthesized compounds. & \url{omdb.mathub.io/} \\
			OQMD & The Open Quantum Materials Database is a database of DFT-calculated thermodynamic and structural properties of materials. & \url{oqmd.org} \\
			Open Materials Database & A database for open access to materials data for research and development. & \url{openmaterialsdb.se} \\
			QM9 Database & A database of small organic molecules for quantum ML models. & \url{quantum-machine.org/datasets/} \\
			ThermoML & An IUPAC standard XML based approach for storage and exchange of experimental and critically evaluated thermophysical and thermochemical property data. & \url{trc.nist.gov/ThermoML} \\
			Thermtest & A database for thermophysical properties of materials. & \url{thermtest.com/thermal-resources/materials-database} \\
			\bottomrule
	\end{tabular}}
	\caption{Different repositories targeted for Machine Learning in Material Science}
	\label{tab:ml_materials_science}
\end{table*}
\begin{table}[h!]
	\centering
	\begin{tabular}{p{3cm}p{7cm}p{4cm}}
		\toprule
		\textbf{Name} & \textbf{Short Description} & \textbf{Website} \\
		\midrule
		CatLearn & An open-source Python library for ML in catalysis. & \url{catlearn.org} \\
		Catalysis-Hub & A platform providing access to data and tools for research in catalysis. & \url{catalysis-hub.org} \\
		Catalysis Knowledge & Provides access to a comprehensive database of catalytic materials and related data. & \url{catalysisknowledge.com} \\
		Catalyst Design & A platform for designing and optimizing catalysts using ML. & \url{catalystdesign.org} \\
		CMR & Catalyst Materials Repository, a database for catalyst materials and their properties. & \url{catalystmaterials.org} \\
		Open Catalyst Project & An open-source project for developing ML models for catalysis. & \url{opencatalystproject.org} \\
		Reaction Mechanism Generator (RMG) & An open-source software for automatic generation of chemical reaction mechanisms. & \url{rmg.mit.edu} \\
		XDL & An open-source language for describing experiments in catalysis and other fields. & \url{xdl.org} \\
		CatApp & Catalysis Applications Database for accessing and sharing catalytic data. & \url{catapp.org} \\
		ChemTS & An AI tool for automated molecular design for catalysis & \url{chemts.kazusa.or.jp} \\
		ChemCatBio & Consortium focused on catalysis for biofuels and bio based chemicals. & \url{chemcatbio.org} \\
		Catalysis Data Science & A platform for data science approaches in catalysis research. & \url{catalysisdatascience.org} \\
		CatDATA & A database for catalyst characterization and performance data & \url{catdata.org} \\
		CatKB & The Catalysis Knowledge Base, a comprehensive resource for catalytic data and information. & \url{catkb.org} \\
		Catalyst Informatics & A platform for integrating informatics tools in catalyst design and optimization. & \url{catalystinformatics.org} \\
		ChemHTPS & High-throughput computational screening platform for catalysis. & \url{chemhtps.org} \\
		\bottomrule
	\end{tabular}
	\caption{Projects on Machine Learning in Catalysis}
	\label{tab:ml_catalysis}
\end{table}
\section{Application of ML}
While the research on CNs and other catalysts are prolific the application of ML in this field is relatively new. ML offers a promising alternative to accelerate the discovery of new materials and can provide pathway to enhance the performance of the traditional photocatalysts. From data curation to model selection and even to the generative AI regime a handful of works on catalysts, especially on CNs are found in literature.

Large datasets are built on experimental and theoretical findings for ML application over time. Most of the online repositories of materials' data and tools for ML application are presented in Table-6. A separate Table-7 is compiled dedicated to the online platforms for catalysts present. Through analysis of material properties, and performance metrics on large dataset, ML algorithms can identify patterns and correlations easily which demands much experience otherwise.

While data curation is the essential first step in ML modelling, ethical considerations are also important. This includes adherence to legal and ethical guidelines on data privacy, version control, and proper documentation for reproducibility and transparency of results.

Throughout the synthesis processes and diverse applications of CNs, the pervasive integration of ML is evident. Liquid-phase exfoliation (LPE) stands as the preeminent technique for synthesizing 2D g$-$C$_3$N$_4$ nanosheets. Key parameters assessing solvent efficacy in LPE include the free energy necessary (G$_{exf}$) to exfoliate a unit area of layered materials into individual sheets and the solvation free energy per unit area of a nanosheet (G$_{sol}$). ML algorithms were employed to predict G$_{exf}$ and G$_{sol}$ for g$-$C$_3$N$_4$, leveraging a database derived from MD simulations encompassing 49 distinct solvents with diverse chemical structures and properties \cite{shahini2023predicting}. Among the six ML methods investigated, the Extra-Tree regressor emerged as the optimal performer.

Band-gap modulation through doping of g$-$C$_3$N$_4$ represents a widely adopted technique. Modelling to predict band-gaps of doped g$-$C$_3$N$_4$ involved the utilization of experimental data derived from 105 g$-$C$_3$N$_4$ based compounds \cite{owolabi2021prediction}. The low correlation coefficient observed between surface area and bandgap in doped g$-$C$_3$N$_4$ indicated that the descriptor and target variables are not linearly correlated. Finally, the bandgap predictions from the SVR model perfectly aligned with experimental data points.
\subsection{Machine Learned Potentials}
One of the active field on theoretical material science is the development of machine-learning interatomic potentials (MLIP). While, classical force fields (e.g., Lennard-Jones) were handcrafted based on physical intuition and experimental data, MLIPs are generated using modelling on large simulated datasets. For example, CHGNet and M3GNet MLIP models are trained on Material Project database \cite{deng2023chgnet,chen2022universal}. Depending on the descriptors, MLIPs can be broadly categorized into two classes:  those that rely on local descriptors and those that use graph-based descriptors. Local descriptors are computationally efficient and scalable, but they may not be able to capture long-range interactions. On the other hand, graph-based descriptors provide higher accuracy by modeling global atomic relationships considering entire structures, however complex they are. However, they are more computationally demanding and harder to interpret due to their complex nature, especially when using deep learning models are employed. A comprehensive description on the development of different MLIPs can be found in \cite{mortazavi2024recent}. From Behler and Parrinello's local descriptors based MLIP to recent developments on high-order tensor message passing interatomic potential have been discussed \cite{behler2007generalized,wang2024n}. Different use of MLIP on catalysis research is compiled by Choung \etal \cite{choung2024rise}.

The utilization of MLIP has significantly advanced computational studies in materials science, particularly in predicting complex properties such as phononic dispersion, thermal conductivity, and mechanical responses in various CN based materials. Shapeev \etal. pioneered the development of moment tensor interatomic potentials using active learning methods under classical or ab-initio molecular dynamics (AIMD) data, enabling accurate calculations of phononic responses, as highlighted in Novikov \etal \cite{novikov2020mlip}. This approach was further applied to investigate the phononic dispersion and stress-strain relationships in C$_6$N$_7$ based 2D monolayers by Mortazavi \etal \cite{mortazavi2022combined}, and to study thermal expansion in complex nanomembranes \cite{mortazavi2022exploring}. The versatility of MLIP extends to various 2D CNs including C$_3$N$_4$, C$_3$N$_5$, and C$_3$N$_6$, demonstrating its effectiveness in understanding the thermal and mechanical behaviors of these materials \cite{mortazavi2022first, mortazavi2020nanoporous}. Moreover, MLIP was successfully employed in non-equilibrium MD simulations on C$_2$N to predict lattice thermal conductivity and mechanical properties, illustrating its capability to simulate dynamic material responses under non-equilibrium conditions as detailed by Arabha \etal \cite{arabha2021thermo}. 

ML force fields have become increasingly prevalent in MD simulations, particularly in applications requiring extended timescales and complex phenomena such as tautomerism in materials like dual defect-modified g-C$_3$N$_4$. Agrawal \etal utilized nonadiabatic MD simulations employing MLP to investigate tautomerism, a phenomenon where molecules exist in equilibrium between different structural isomers, which is challenging to capture with traditional AIMD methods limited to short timescales (picoseconds). This approach leverages MLPs to approximate interatomic potentials more accurately over longer simulation times, allowing researchers to study dynamic processes like tautomerism in detail. The study exemplifies the utility of ML based approaches in overcoming the temporal and computational limitations of AIMD, thereby advancing our understanding of complex molecular behaviors in photocatalytic materials \cite{agrawal2024photocatalytic}.

Not just for crystallised structures, machine learned potentials are useful on amorphous systems as well. Jeong \etal comprised total 64 atoms' configurations of Carbon and Nitrogens for AIMD of different densities (2.0, 2.45, 2.95, 3.2, and 3.5 gm cm$^{-3}$) \cite{jeong2024integrating}. Using 16 radial and 54 angular components symmetry functions as input features a neural network potential (NNP) for amorphous CN was trained using the SIMPLE-NN package \cite{lee2019simple}. A comprehensive database for 216 atoms was constructed on a variety of amorphous CN structures, those investigated for spectral fingerprints using DFT approach. 


\begin{figure*}[t!]
	\centering
	\includegraphics[width=0.8\textwidth]{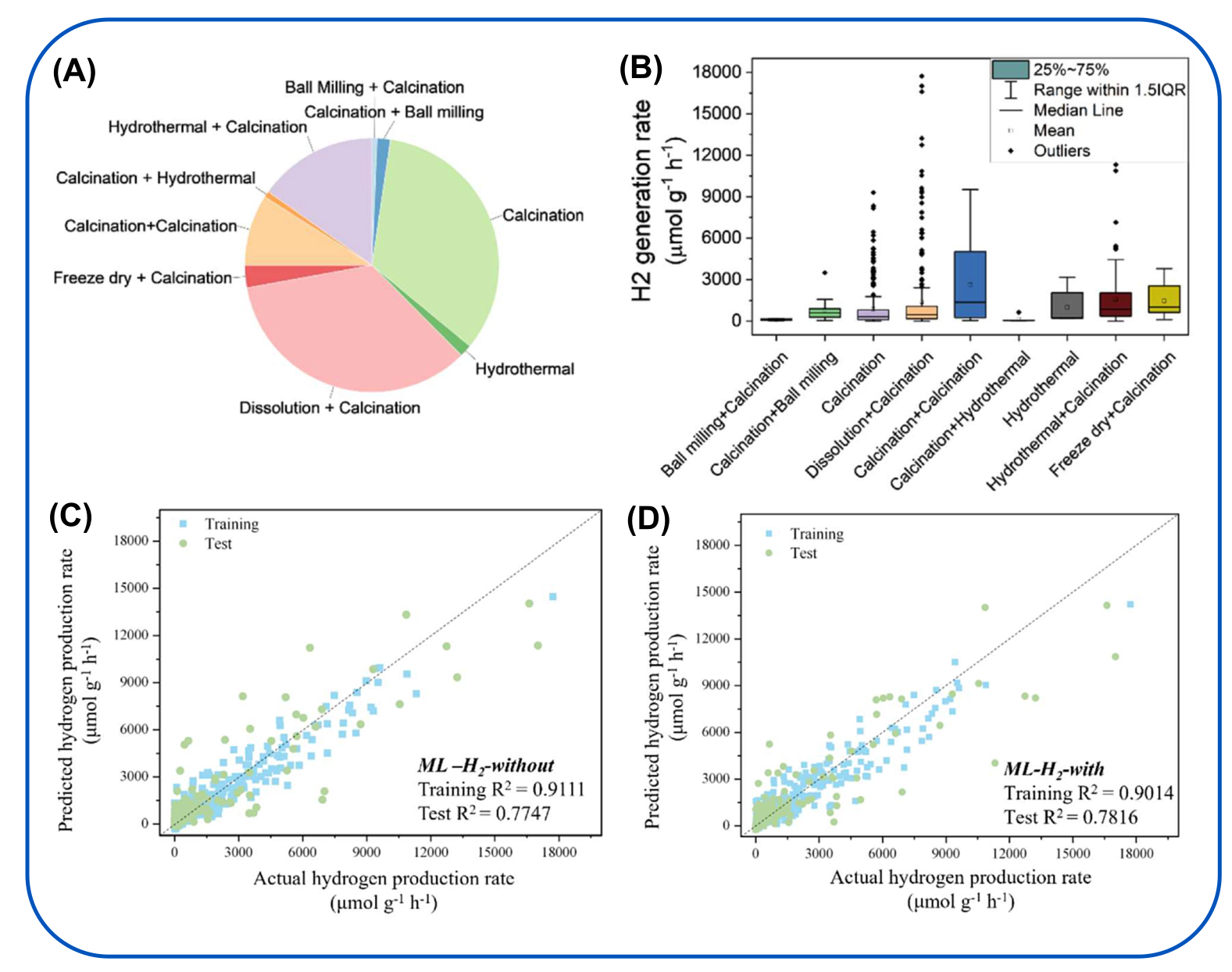}
	\caption{(A) Frequency distribution of each synthesis method within the datasets. (B) H$_2$ production rates categorized by treatment methods. (C) and (D) Machine learning model performance metrics for predicting H$_2$ production rates without and with considering bandgap and specific surface area in the input data, respectively. Reprinted with permission from ref \cite{yan2022development}. Copyright 2022, Elsevier.}
	\label{fig:HER}
\end{figure*}
\begin{figure*}[t!]
	\centering
	\includegraphics[width=0.8\textwidth]{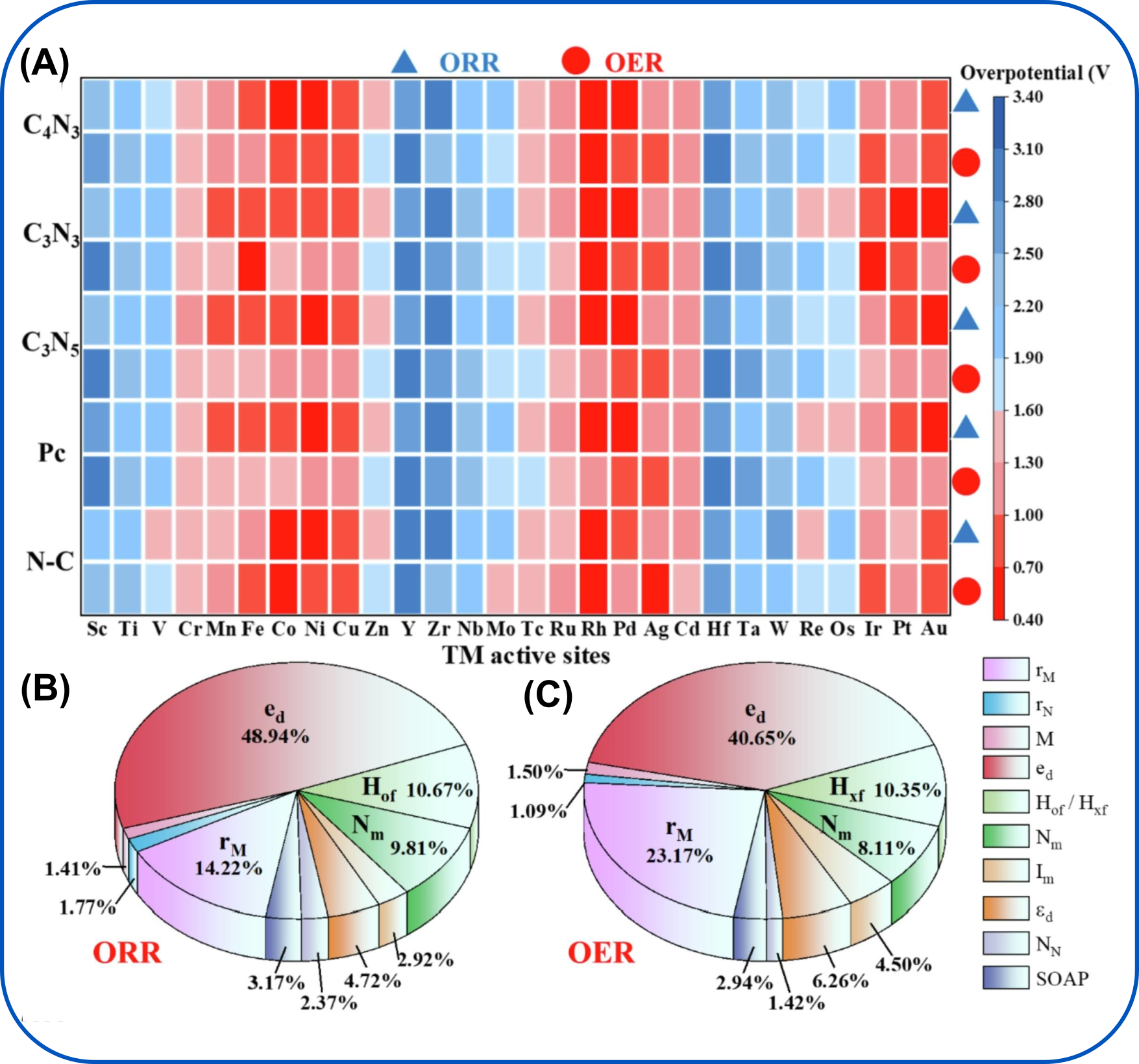}
	\caption{(A) Heat map of ML-predicted ORR/OER overpotentials for SAC$@$CNs (C$_4$N$_3$, C$_3$N$_3$, C$_3$N$_5$, Pc, N-C). Redder colors indicate higher catalytic activity. Blue triangles represent ORR, red circles represent OER. (B) and (C) Feature importance in RFR models for ORR and OER prediction, respectively. Reprinted with permission from ref \cite{wan2022revealing}. Copyright 2022, Elsevier.}
	\label{fig:OER}
\end{figure*}
\subsection{Catalysis Research}
Discovery of CNs and their derivatives are regarded as a revolutionary leap toward artificial catalysis. Catalytic activity involves two half-reactions: (i) Hydrogen Evolution Reaction (HER) at cathode, and, (ii) Oxygen Evolution / Reduction reaction (OER/ORR) at anode. In photocatalysis, photochemical reactions are driven by photons with energies greater than the bandgap. These photons generate electron-hole combination, If they do not recombine, then HER and OER reactions are initiated at the conduction band (CB), and valence band (VB) by electrons and holes, respectively. For that certain conditions have to be satisfied. For example, in case of water splitting the maximum of VB should be more positive than the OER potential ($E_{H_2O/O_2}$=1.23, 0.81 V for pH=0, 7), and the minimum of CB should be more negative than the HER potential ($E_{H^+/H_2}$=0, $-0.41$ V for pH=0, 7 in normal hydrogen electrode (NHE) scale): 
\begin{eqnarray*}
	H_2O+2h^+ &\rightarrow & 2H^+ + \frac{1}{2}O_2 \nonumber\\
	2H^+ + 2 e^- &\rightarrow & H_2  \quad  
\end{eqnarray*}

Both of the HER and OER generally are not single step processes. The calculation of Gibbs free energy change for each sub-step is essential for finding the reactional pathway. This can be expressed as

$$	\Delta G = \Delta E + \Delta ZPE - T\Delta S + \Delta G_U + \Delta G_{pH}$$

where, $\Delta E$, $\Delta ZPE$, $\Delta S$, $\Delta G_U$, and $\Delta G_{pH}$ represent the change of adsorption energy, zero-point energy (ZPE), entropy (S), change of free energy contribution due to electrode potential and ion concentrations, respectively \cite{yuan2022c6n3}. For HER/OER, $\Delta E$ can be computed using DFT through total energy calculations on catalysts with and without adsorbed Hydrogen/Oxygen atom and isolated H$_2/$O$_2$ gas molecule as:

$$\Delta E = E_{H+surface} - E_{surface} - \frac{1}{2} E_{H_2/O_2}$$

\subsubsection{Hydrogen Evolution Reaction}
The Hydrogen Evolution Reaction  is a fundamental electrochemical process that involves the reduction of protons (H$^+$) and electrons (e$^-$) to produce hydrogen gas. This reaction plays a crucial role in various energy-related technologies, particularly in hydrogen production and fuel cell applications.

Aiming to fabricate single atom catalysts (SACs) from 3d, 4d, and 5d transition metals (TM), metal atoms were set as the active sites on C$_6$N$_3$ sheets by Yuan \etal \cite{yuan2022c6n3}. As a primary step, the electronic properties of TM such as atomic number and charge, number of valence electrons, number of d electrons and positions of d band centres, atomic radius, electronegativity in Pauling scale, first ionization energy, electron affinity were considered as features. However, the result was not satisfactory, hence, the distance between TM and N atoms and the atomic charge of the N atom were added as features. Physical parameters of TM showed low linear correlation with the local structure around TM. It demonstrate the importance of feature engineering as preprocessing step.

Similar work using DFT and ML on SAC$@$C$_3$N$_4$ was done bu Umer \etal \cite{umer2022machine}. To estimate the model performance, the Monte-Carlo cross validation method was employed. Among the nine distinct ML regression models tested for predicting HER activities, the CatBoost (an extension of GBR) regression model demonstrated the highest accuracy.

Often, a comparison of different models on same data is beneficial. The final model selection is depends on achieving the optimal balance among different factors such as predictive accuracy, computational efficiency, interpretability, and scalability of ML models. They compared different ML regression algorithms including LR, RR, RF regression, feed-forward ANN regression, and finally, found GBR as best performing method to predict HER activity for TM$@$C$_6$N$_3$.

Another study on SACs on a different CN, g$-$C$_3$N$_4$ was conducted using DFT data by Jyothirmai \etal. They identified B, Mn, and Co embedded on g$-$C$_3$N$_4$ as highly competent catalysts for hydrogen production from a series of different metals and nonmetals \cite{jyothirmai2023accelerating}. Comparing different regression models, SVR was found as most effective in this particular study.

In material science most of the dataset are often not as large as needed for effective model building. Hence, multiple datasets or even different types of data have to be integrated to create a comprehensive dataset. A database totalling 767 instances was built from of 106 research articles on experimental g$-$C$_3$N$_4$ doped by different elements for photocatalytic hydrogen production by Yan \etal \cite{yan2022development}. Along with H$_2$ production criteria, material properties and synthesis conditions were fed to the ML models, and the Hydrogen production rate was derived as output (see, Figure--\ref{fig:HER}(A--D)). Instead of doing separate feature selection, RF algorithms was adopted for its intrinsic property of rewarding higher weigh to important features. Data cleaning is an essential process to ensures data quality and consistency. Eight encoding methods and five scaling techniques were used to scale numerical variables. Also to handle missing values, tree based ML algorithms like XGBoost and CatBoost were selected. Bayesian optimization was adopted for tuning the hyperparameters of ML model. Finally a perfect synthesis condition including the choice of precursor and dopant was extracted.

Besides SACs, heterostructures of CNs with other materials for catalysis steals the show iven in ML regime.  In a work on a g$-$C$_3$N$_4/$MX$_2$ heterostructure, another popular scaling method of numerical features “MinMaxScaler”  was utilised from the Scikit-Learn library in a hunt for efficient HER \cite{jyothirmai2024machine}.

\subsubsection{Oxygen Evolution / Reduction reaction}
The oxygen evolution reaction and oxygen reduction reaction are fundamental electrochemical processes that involve the exchange of oxygen atoms and electrons at the surface of electrodes. OER refers to the process where oxygen molecules are split into oxygen ions and electrons. ORR, on the other hand, involves the reduction of oxygen molecules by accepting electrons. Both reactions are pivotal in energy conversion and storage technologies, influencing the efficiency and performance of various electrochemical devices.

To validate the feasibility of a single TM embedded in defective g$-$C$_3$N$_4$ for bifunctional oxygen electrocatalysis, a ML model was employed to elucidate the intrinsic correlation between targeted adsorption energy and descriptors \cite{niu2021single}. The input dataset included 10 descriptors pertaining to structural and atomic properties. Given the limited size of the dataset, the GBR model was selected as the ML algorithm, which exhibited exceptional performance.

Using DFT, ML, and a cross validation scheme, Wan \etal established the best performing ML regression models for ORR/OER \cite{wan2022revealing}. Those models effectively characterized the relationships between readily accessible physical and chemical properties and the overpotentials of C$_2$N, C$_3$N, and C$_3$N$_4$ related SACs (Figure--\ref{fig:OER}(A--C)). Through this approach, three promising oxygen electrocatalysts were identified with low overpotentials.

In catalytic research stability of the material is a prime concern. The experimental data for stability comes in a time domain data form. To forecast the stability of prepared samples, data from chronoamperometric measurement for 800~s was taken \cite{koyale2024synergistic}. Long and short term
memory (LSTM) technique of recurrent neural network was utilised. The model forecasted  a nominal 0.12\% decrease in stability of photo-anodes for an extended period.
\begin{figure*}[t!]
	\centering
	\includegraphics[width=\textwidth]{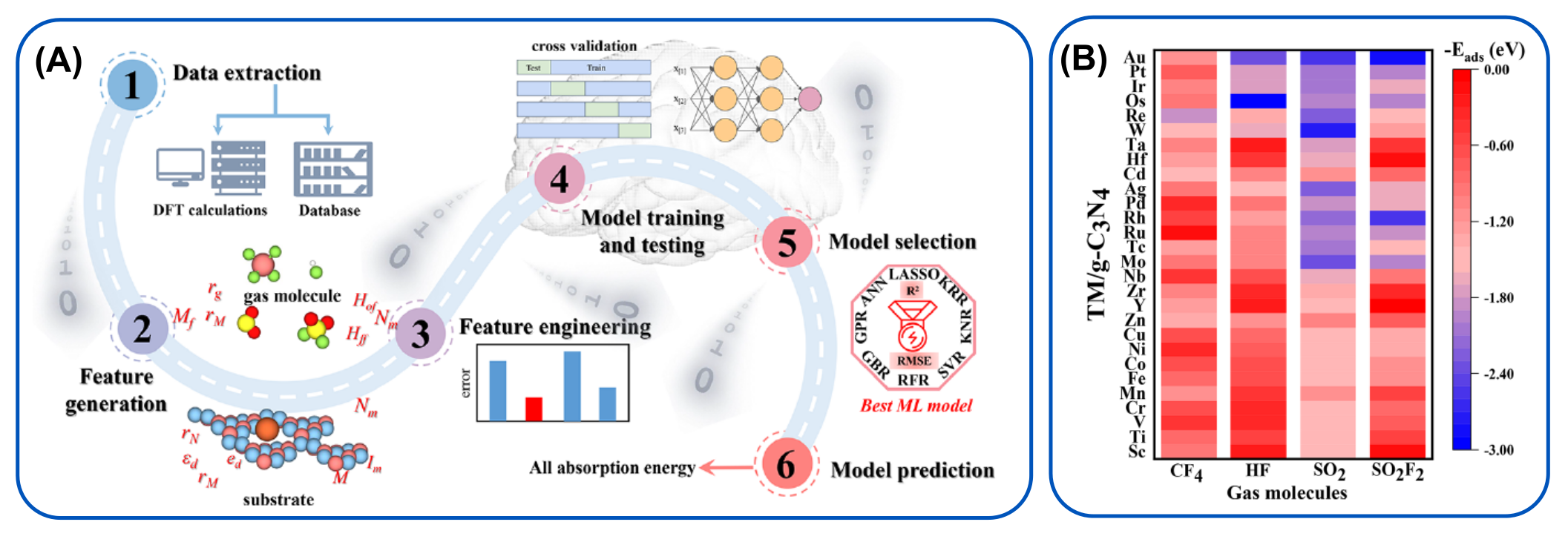}
	\caption{(A) Machine learning workflow for exploring interaction strengths among four decomposed CF$_3$SO$_2$F gas products and 28 TM$/$g$-$C$_3$N$_4$ combinations, employing common ML algorithms. (B) Absorption energies calculated for interactions between 28 distinct types of TM supported on g$-$C$_3$N$_4$ and four different decomposition gases. Reprinted with permission from ref \cite{wan2023high}.	Copyright 2023, American Chemical Society.}
	\label{fig:gas-sensor}
\end{figure*}

\subsection{Environmental Remedy}
Graphitic CN has established itself as a versatile material with wide-ranging applications in environmental remediation and sustainable technology. Its two-dimensional structure, enriched with nitrogen, enables efficient photocatalytic activity under visible light, making it effective in degrading organic pollutants and oxidizing harmful gases in both water and air. This capability finds practical use in wastewater treatment, air purification systems, and industrial emissions control, where g$-$C$_3$N$_4$'s ability to be tailored for enhanced adsorption further enhances its utility. 

One of its successful application is found in CO$_2$ reduction reaction (CO$_2$RR). The CO$_2$RR involves the electrochemical reduction of carbon dioxide (CO$_2$) to produce valuable chemicals or fuels. It typically occurs in an electrolytic cell where CO$_2$ is supplied to the cathode, and a voltage is applied to drive the reduction process. Beside carbon monoxide (CO), CO$_2$RR can also produce other products such as methane (CH$_4$), ethylene (C$_2$H$_4$), formic acid (HCOOH), and even higher hydrocarbons, depending on the catalyst and operating conditions used.	CO$_2$RR research aims to improve the efficiency and selectivity of these reactions to make them commercially viable for carbon capture and utilization, as well as for renewable energy storage applications.

Zhu \etal used DFT and ML together to predict free energy change in the reactional pathway for reduction of CO$_2$ by metal-oxide-frameworks (MOF), metal-zeolites, metal-doped 2D materials including graphene, and g$-$C$_3$N$_4$ \cite{zhu2022machine}. The descriptors include the reaction pathways, the metal involved, the charge transfer (CT) occurring between the metal and reaction intermediate, the hydrogen bond interaction between the intermediate and the zeolite framework, and the geometric configuration.  A variety of ML algorithms were tested including GBR, MLR, XGBoost, SVR, LASSO, Extra-Trees, KNN, Decision tree, LR, and least absolute shrinkage, ANNs. A total of 34 catalysts were tested and predicted the final products in CO$_2$RR and with 91\% accuracy. XGBoost, ExtraTrees, and GBR performed best.

Goliaei evaluated cobalt single atoms and Co$_2$O$_2$ clusters supported on g$-$C$_3$N$_4$ as catalysts for CO$_2$RR \cite{moharramzadeh2024photocatalytic}. Linear regression, neural-network based, and ensemble tree model combined with GBR were considered. After training these models on a designated training set, they assessed their performance by comparing RMSE on a test set. The ensemble tree model which was proved to be useful for the studies on energy materials, exhibited superior performance, highlighting its effectiveness for small datasets in this application. Employment of ML in CO$_2$RR by CNs well-documented across various studies \cite{xiang2022solving}.

The study by Zhao \etal investigated the efficacy of single metal atom catalysts embedded on g$-$C$_3$N$_4$ for formic acid dehydrogenation, focusing on their impact on the adsorption energy of formic acid. According to the results from ML models, the adsorption strength of formic acid on M$@$g$-$C$_3$N$_4$ is predominantly influenced by two critical features: the higher electronegativity of the metal atom and the more negative d-band centre of the metal atom. The ML scheme utilized several widely adopted models, including RR, RFR, ANN, and GBR, implemented through the scikit-learn package. These models were employed to analyse and predict data patterns, leveraging their respective strengths in regression tasks to achieve robust and accurate predictions within the study's framework.

Chen \etal targeted nitrogen reduction reaction (NRR) by TM single atom photocatalysts embedded on C$_3$N$_4$ from theoretical data \cite{chen2024first}. Expressing $\Delta G_U = eU$, where, $U$ is the applied electrode potential, they set the onset potential U as the target property. They considered both of electronic and geometric properties as features, totalling 8 in number. Sure Independence Screening and Sparsity Optimization (SISSO) is a ML method curated for materials science applications \cite{ouyang2018sisso}. It aims to identify as well as optimize the most relevant descriptors while maintaining interpretability side-by-side. Using backward elimination method and SISSO, they identified three descriptors correlating with the target property. Finally Ru was predicted as a promising candidate singe photocatalyst residing on g$-$C$_3$N$_4$ for NRR. Tungsten atoms supported on g$-$C$_3$N$_4$ was identified as promising SACs for the NRR, characterized by low limiting--potential and high selectivity towards ammonia, as predicted by ML models by Zhang \etal \cite{zhang2023self}. Similar study on NRR by TM embedded C$_6$N$_6$ was conducted by Mukherjee \etal \cite{mukherjee2022performance}.

Li \etal employed RF, XGBoost and GBR models to predict the photocatalytic purification rate of NO on g$-$C$_3$N$_4$ based catalysts using 255 data points with 14 input features \cite{li2024prediction}. They employed Shapley additive explanation, feature importance analysis, and partial dependence plots to investigate the influence of input features on the target variable, the NO removal rate.

Miodyńska \etal take the advantage of ML for selecting optimal  Cs$_3$Bi$_2$X$_9$ stoichiometry with various amounts of X $=$Cl, Br, and I halides and their combinations on C$_3$N$_4$ matrix for degradation of pollutants under visible light \cite{miodynska2022lead}. PCA was used for virtual screening of a plethora of binary structures. Based on the structural similarity grouping of the Cs$_3$Bi$_2$X$_9$ perovskites were done and a possible relationship was determined between these structures and their photocatalytic performance.
In a separate investigation, ANN based ML modeling was utilised to estimate the photodegradation efficiency of g$-$C$_3$N$_4$ for the degradation of the organic pollutant Reactive Black 5 \cite{dorraji2017photocatalytic}. 

Antibiotic pollution in water is another growing environmental concern due to the widespread use of antibiotics in healthcare, agriculture, and veterinary practices. Once in aquatic environments, antibiotics can persist and accumulate, posing risks to aquatic organisms and potentially entering the food chain. Gordanshekan \etal synthesised Bi$_2$WO$_6/$g$-$C$_3$N$_4$ and Bi$_2$WO$_6/$TiO$_2$ heterojunctions for adsorption/degradation of Cefixime antibiotic in aquas environment \cite{gordanshekan2023comprehensive}. ML scheme involving ANNs with 18 hidden neurons performed best for predicting of the kinetics of the photocatalytic reaction. In another study on g$-$C$_3$N$_4/$Ce$-$ZnO$/$Ti, the Cefixime degredation was taken as the targeted output for ANN and the input parameters were number of electrode, pH, power of visible light, electrolyte concentration, pollutant concentration \cite{sheydaei2021visible}.
\begin{figure*}[t!]
	\centering
	\includegraphics[width=0.7\textwidth]{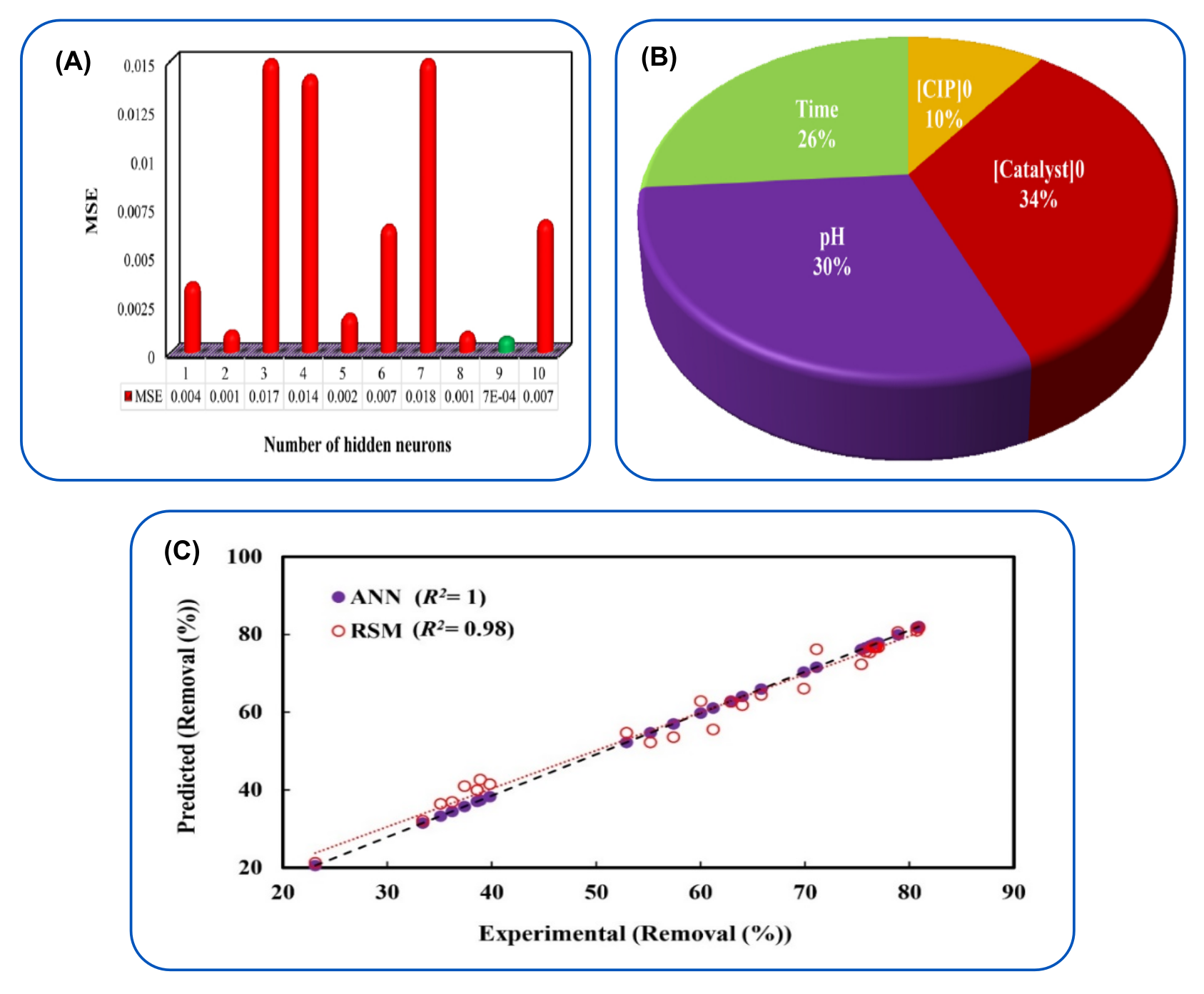}
	\caption{(A) Mean squared error (MSE) of designed networks plotted against the number of hidden neurons. (B) Relative importance of operational parameters on the efficiency of CIP (Ciprofloxacin) removal. (C) Comparison between experimental results and predictions obtained using Response Surface Methodology coupled with ANN. Reprinted with permission from ref \cite{qarajehdaghi2023quaternary} Copyright 2023 Elsevier.}
	\label{fig:antibiotic}
\end{figure*}
Eight ML methods were employed to predict the performance of g$-$C$_3$N$_4$ based photocatalysts in degrading tetracycline \cite{song2024prediction}. Starting from 431 experimentally obtained records from literature, the XGBoost model utilized an optimized leaves parallel processing strategy to effectively capture the experimental parameters of tetracycline photodegradation. This approach mirrors the reaction generation process involved in tetracycline degradation facilitated by g$-$C$_3$N$_4$ based photocatalysts.

Multivariate ANNs were used for predicting the Ciprofloxacin degradation efficiency of  ZnO$/$g$-$C$_3$N$_4$ by Gupta \etal \cite{gupta2021multivariate}. Another study investigated the efficiency of a quaternary composite and assessed the relative importance of operational parameters using ANN \cite{qarajehdaghi2023quaternary}. The importance of operational environments as features and the performance of ANN is presented in Figure--\ref{fig:antibiotic}. 
The photocatalytic degradation efficiency of tetracycline hydrochloride by Fe$_3$O$_4/$g$-$C$_3$N$_4/$rGO predicted using ANN hybridised with Particle Swarm Optimization (PSO) model was found to be at par with experimental results by Shan \etal \cite{shan2023photocatalytic}.

Not only the removal of pollutant, environmental research also aims the production of pollution free energy sources, like green hydrogen production, synthesis of biodiesel, enhancing the performance of solar cells, etc. Taib's research focused on transesterification of algal oil for biodiesel production, aiming to advance pollution-free energy sources \cite{taib2024rsm}. The study utilized ANN and GRNN for modeling and optimizing the transesterification process. The methanol to oil ratio is identified as a critical factor influencing the yield of the blend in transesterification reactions.

A recent study involved the adsorption of gases produced by decomposition of CF$_3$SO$_2$ (e.g., SO$_2$, SO$_2$F$_2$, and CF$_4$)  on TM$@$g$-$C$_3$N$_4$ as presented in Figure--\ref{fig:gas-sensor}. The focus was on predicting adsorption energies with high accuracy using an ML approach. A comprehensive process involving feature engineering, data extraction, and model training was undertaken with Scikit-learn and PyTorch packages. Eight supervised ML algorithms were employed: GBR, RF, SVR, KNR, KRR, LASSO, ANN, and GPR. Among these, the SVR model demonstrated superior performance on both training and test sets, achieving a lower $RMSE = 0.223 ~eV$ and a higher score of $R^2 = 0.911$, indicating its effectiveness \cite{wan2023high}.
\begin{figure*}[t!]
	\centering
	\includegraphics[width=\textwidth]{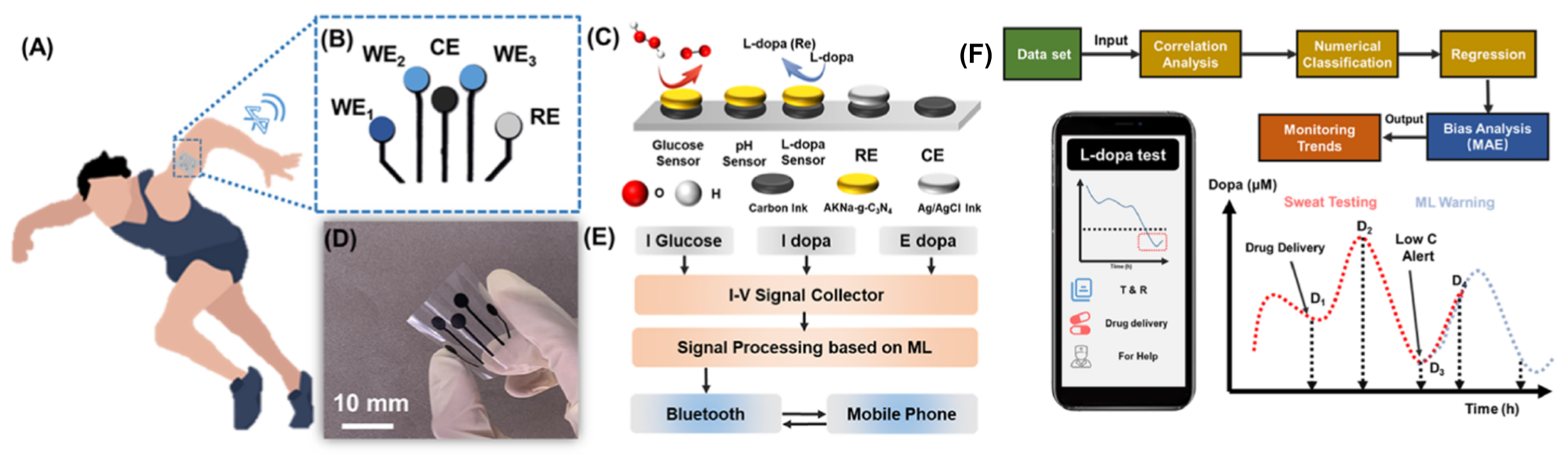}
	\caption{A comprehensive workflow of wireless sweat sensing devices combining (A) Schematic diagram for real-time monitoring, (B) primary structural components within the sensing chip, (C)A model illustrating the operational mechanism of electrode, (D) Photographic depiction of an unencapsulated sensing chip, (E) Logical diagram delineating the operational sequence of the wireless transmission module, focusing on signal acquisition, transduction, and data output from three distinct sensors and, (F) Schematic diagram illustrating the processing model for simulated human drug delivery based on ANN. Reprinted with permission from ref \cite{yu2022artificial} Copyright 2022 American Chemical Society.}
	\label{fig:dopamine}
\end{figure*}
\subsection{Bio-medical application}
Material science plays a pivotal role in medical diagnostic technologies through enabling the evolution of advanced materials that improve sensitivity, and reliability in identification tools. Across diverse fields such as biosensors, diagnostic imaging, lab-on-a-chip devices, and nanomedicine, material scientists are leaving their footprints.

Dopamine, a key neurotransmitter in humans and mammals, regulates central nervous system functions including behaviour, learning, and memory. Dysregulation of dopamine levels is linked to neurological disorders like Parkinson’s disease, schizophrenia, and Tourette syndrome, underscoring the critical need for accurate and sensitive dopamine detection in clinical diagnosis. Electrochemiluminescence (ECL) is an effective tool for dopamine detection and C$_3$N$_4$ derived materials are promising agent for ECL detection. Li \etal carried out multiparameter concentration prediction of dopamine using DNN assisted technique on Ag@g--C$_3$N$_4$ \cite{li2024electrochemiluminescence}. Each of the eight dopamine concentrations underwent five measurements, resulting in a data matrix of 16 parameters across 8 concentrations and 5 parallel tests, which was then analyzed using the DNN algorithm. There was a strong correlation between experimental and ML predicted dopamine concentrations across the range from 0.1 nM to 1 mM, demonstrating the accurate prediction capability of the multicolour ECL detection array in conjunction with the DNN algorithm. The DNN-assisted model achieves a high $R^2 = 0.99861$ and $RMSE = 0.0997$ value, indicating excellent predictive accuracy for dopamine concentration.

In Parkinson’s disease treatment, monitoring L dopa levels in sweat offers insights into pharmacological management and non-motor complications. Nanozymes, like layered g$-$C$_3$N$_4$ are promising due to their robust catalytic activity, stability, and customizable performance, potentially replacing natural enzymes \cite{zhu2019ultrathin, zhang2019modified}. Customized dose criteria based on long term serial assessments of L dopa levels in sweat could optimize patient treatment. Utilising raw electrical signals of sweat analytes, along with factors like time, pH, and daily activity, as training data for ML models, enabled real time prediction of L-dopa levels (Figure--\ref{fig:dopamine}) in a recent study by Yu \etal \cite{yu2022artificial}. This approach supported more effective pharmacological management of Parkinson’s disease patients. Various ML algorithms including LR, MLRs, CNNs, and ANNs were employed to predict trends in L-dopa levels in sweat post individual dosing. These models also provided estimations for the optimal timing of next doses based on insights gained from the dataset.

In a significant work, Au@g--C$_3$N$_4/$ZnO was synthesized using a straightforward chemical technique to advance the development of triboelectric nanogenerators (TENG) for achieving high accuracy in biomechanical motion recognition \cite{das2023flexible}. The outputs from ZnAuCN based TENG sensors were analysed using supervised ML. Notably, no significant performance improvement was observed with alternative algorithms such as ANN or SVM.



\section{Future Perspective of ML Research in GAI Regime}
CNs are regarded as future materials with prospects in advanced application in versatile fields like photocatalysis, energy storage, and electronic devices. The integration of ML in the research of CN materials holds immense potential to revolutionize the field. ML-driven approaches significantly advance the development and application of CNs as discussed in the previous sections by accelerating discovery, optimizing synthesis, predicting properties, and enabling sustainable practices.

Generative AI is the next paradigm shift in the field of automation. In its early stages, it is showing promise in material science.  The Deep Inorganic Material Generator (DING) proposed by Pathak \etal uses conditional VARs (CVAEs) and a predictor module consisting of three DNNs trained to predict the formation enthalpy, the volume per atom and the energy per atom to discover new materials \cite{pathak2020deep}. Using materials from the ICSD database as training data, a GAN-based ML model, MatGAN for the efficient generation of new hypothetical inorganic materials \cite{dan2020generative}. Using this Song \etal discovered 267,489 new potential 2D material compositions, with 1,485 highly probable compositions, predicted 101 crystal structures, and confirmed 92 2D/layered materials via DFT formation energy calculations \cite{song2021computational}.

Court \etal propose an autoencoder-based generative deep representation learning pipeline for optimizing 3D crystal structures and predicting eight target properties: formation energy per atom, energy per atom, bulk modulus, shear modulus, refractive index, dielectric constant, Poisson ratio, and bandgap \cite{court20203}. Based on data from the Materials Project, the pipeline generated novel binary alloys, ternary perovskites, and Heusler compounds using VAE and predicted the properties using a GNN. A plethora of recent generative AI approaches for automated material discovery have focused on various applications, including drug discovery, solar cells, and battery materials, ranging from novel molecules to complex crystal structures \cite{zhao2021high,gomez2018automatic,hu2019matganip,joshi2019machine,merchant2023scaling}. Some recent review articles have compiled various works in this field \cite{park2024has,menon2022generative}.

Not only limited to material discovery, but GAI models have also been employed for the autonomous determination of phase diagrams with minimal human supervision \cite{arnold2024mapping}.

In catalysis research, GAI is gaining popularity. The GAN based approach, called Catalysis Clustering, incorporates domain knowledge into the clustering process by generating catalysts—special synthetic points drawn from the original data distribution. These catalysts are verified to improve clustering quality based on a domain-specific metric \cite{andreeva2020catalysis}. DFT was combined with a GAN to artificially propose heterogeneous catalysts using a DFT calculated dataset. This approach was demonstrated on the NH$_3$ formation reaction on Rh--Ru alloy surfaces \cite{ishikawa2022heterogeneous}. Conditional GAN predicted transition state (TS) geometries by mapping reactants and products directly to the TS space. This method generated TS guesses from Cartesian coordinates alone, bypassing costly reaction path mapping and aiming for simplicity, general applicability, and easy expansion \cite{makos2021generative}.

GAI is targeted to make the coding or research job easier in every sense.  The Generative Toolkit for Scientific Discovery (GT4SD) is an open-source library for training and using state-of-the-art generative models in organic material design. It supports execution, training, fine-tuning, and deployment via Python or CLI, with pre-trained models also accessible through web apps \cite{manica2023accelerating}.

Our discussion clearly indicates that research on CN-based materials is undergoing a significant paradigm shift. Traditional experimental and computational investigations are now increasingly complemented by ML methods. These methods are proving promising in discovering new CN-supported SACs, CN based heterostructures, and pathways for HER, OER/ORR, and pollutant removal reactions.

Although the application of state-of-the-art generative modeling to CNs has not yet been reported, future ML research is expected to focus on advanced material discovery using generative models, such as GANs and VAEs, to predict novel structures with tailored properties. This research will optimize synthesis pathways and experimental parameters to achieve higher purity and yield, and predict critical properties, including electronic, mechanical, and thermal characteristics. Efforts will also include data augmentation to enhance model training and inverse design to work backward from desired properties to potential structures. Integrating ML with traditional research methods will drive innovation in CN materials, contributing to energy, environmental, and technological advancements for the betterment of society.
\paragraph{\textbf{Declaration of competing interest}}
The authors declare that they have no known competing financial interests or personal relationships that could have appeared to influence the work reported in this paper.
\paragraph{\textbf{CRediT authorship contribution statement}}
\textbf{DM}: Conceptualisation, Editing, Writing original draft. \textbf{SD}: Conceptualisation, Editing, Supervision, Writing original draft.  \textbf{DJ}: Conceptualization, Editing, Supervision, Finalizing Manuscript.
\section*{Acknowledgements}
DM would like to thank the Council of Scientific and  Industrial Research (CSIR), India for providing financial support as Senior Research Fellow. 

\bibliographystyle{elsarticle-num}
\bibliography{ref.bib}
\end{document}